\documentclass[english,onecolumn]{IEEEtran}
\usepackage[T1]{fontenc}
\usepackage[latin9]{inputenc}
\usepackage{amsmath}
\usepackage{amsthm}
\usepackage{amssymb}
\usepackage{graphicx}
\usepackage{esint}

\makeatletter
\theoremstyle{plain}
\newtheorem{thm}{\protect\theoremname}
\theoremstyle{plain}
\newtheorem{cor}[thm]{\protect\corollaryname}
\theoremstyle{definition}
\newtheorem{defn}[thm]{\protect\definitionname}
\theoremstyle{plain}
\newtheorem{lem}[thm]{\protect\lemmaname}

\usepackage{mathrsfs}
\usepackage{algorithm,algpseudocode}

\usepackage{quiver}

\usepackage{colortbl}
\definecolor{lightgray}{rgb}{0.9,0.9,0.9}
\definecolor{lightred}{rgb}{1,0.8,0.8}
\definecolor{lightgreen}{rgb}{0.6,1,0.6}
\definecolor{lightyellow}{rgb}{1,1,0.5}
\definecolor{lightgrey}{rgb}{0.8,0.8,0.8}

\allowdisplaybreaks[1]

\makeatother

\usepackage{babel}
\providecommand{\corollaryname}{Corollary}
\providecommand{\definitionname}{Definition}
\providecommand{\lemmaname}{Lemma}
\providecommand{\theoremname}{Theorem}

\begin{document}
\title{A Pair of Bayesian Network Structures has Undecidable Conditional
Independencies}
\author{Cheuk Ting Li\\
Department of Information Engineering, The Chinese University of Hong
Kong, Hong Kong, China\\
Email: ctli@ie.cuhk.edu.hk}
\maketitle
\begin{abstract}
Given a Bayesian network structure (directed acyclic graph), the celebrated
d-separation algorithm efficiently determines whether the network
structure implies a given conditional independence relation. We show
that this changes drastically when we consider two Bayesian network
structures instead. It is undecidable to determine whether two given
network structures imply a given conditional independency, that is,
whether every collection of random variables satisfying both network
structures must also satisfy the conditional independency. Although
the approximate combination of two Bayesian networks is a well-studied
topic, our result shows that it is fundamentally impossible to accurately
combine the knowledge of two Bayesian network structures, in the sense
that no algorithm can tell what conditional independencies are implied
by the two network structures. We can also explicitly construct two
Bayesian network structures, such that whether they imply a certain
conditional independency is unprovable in the ZFC set theory, assuming
ZFC is consistent.
\end{abstract}

\medskip{}

\section{Introduction}

Bayesian network is a popular graphical model which describes the
dependency between a collection of random variables by a directed
acyclic graph. Given a Bayesian network structure (directed acyclic
graph), the celebrated $d$\emph{-separation} algorithm \cite{pearl1988probabilistic,verma1990causal,geiger1990logic,geiger1990d}
efficiently determines whether a conditional independency (CI) among
the random variables is implied by the network. The simplicity and
algorithmic efficiency of conditional independence inference is one
of the major advantages of Bayesian networks.

When there are multiple Bayesian networks, each representing the knowledge
of one agent, the knowledge of the networks should be combined to
obtain the overall knowledge \cite{matzkevich1992topological,matzkevich1993some,xiang1998verification,pennock1999graphical,wong2001constructing}.
There are two lines of existing works on combining Bayesian networks:
\emph{qualitative combination} about combining the structures of two
Bayesian networks (i.e., the directed acyclic graphs) \cite{matzkevich1992topological,matzkevich1993some,ron1997combining,xiang1998verification,wong2001constructing,del2003qualitative,jiang2005pgmc,li2008recovering,feng2014novel,tabar2018novel},
and \emph{quantitative combination} about combining the structures
and parameters (i.e., the probabilities) of two Bayesian networks
\cite{pennock1999graphical,jiang2005pgmc}. We focus on qualitative
combination in this paper, though we will not restrict the result
to be a combined Bayesian network. Instead, we ask the following more
basic question: 

\smallskip{}

\emph{Given two Bayesian network structures, what are the conditional
independencies implied by them? In other words, what is the set of
conditional independencies that must be satisfied if we know that
the random variables satisfy }both\emph{ networks?}

\smallskip{}

\begin{figure}
\begin{centering}
\includegraphics[scale=0.9]{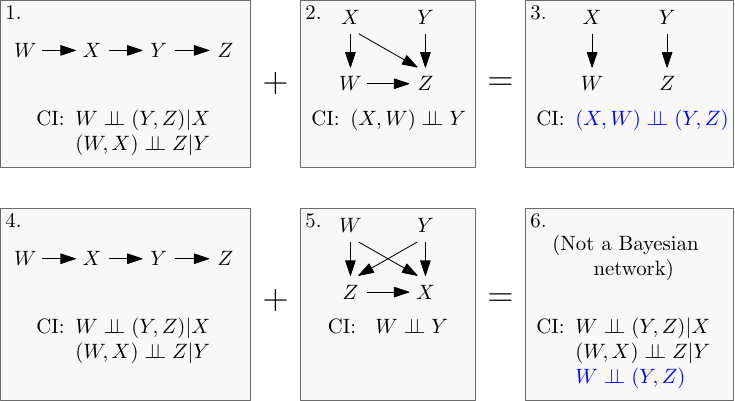}
\par\end{centering}
\caption{\label{fig:example}Two examples of the problem of combining the knowledge
of two Bayesian network structures. In the first example on top, if
we know that the random variables $W,X,Y,Z$ satisfy Network 1 (which
implies the CIs $W\perp\!\!\!\perp(Y,Z)|X$ and $(W,X)\perp\!\!\!\perp Z|X$)
and Network 2 (which implies the CI $(X,W)\perp\!\!\!\perp Y$), then
this is equivalent to $W,X,Y,Z$ satisfying Network 3 (which implies
the CI $(X,W)\perp\!\!\!\perp(Y,Z)$). This CI implied by Network
3 is not implied by Network 1 or 2, and hence we cannot simply take
the union of the set of CIs implied by Network 1 and the set of CIs
implied by Network 2 to obtain the combined set of CIs. The combination
might not even be a Bayesian network, as shown in the second example
below. These observations have been noted in \cite{del2003qualitative}.
This paper shows that generally, the combined set of CIs cannot even
be computed.}
\end{figure}

Refer to Figure \ref{fig:example} for examples. Although this question
is completely solved by $d$-separation \cite{verma1990causal,geiger1990logic}
when we have only one Bayesian network structure, the situation becomes
significantly more complex when we have two networks. This problem
is referred as ``union of independence models'' in \cite{del2003qualitative},
though the set of CIs implied by the two networks together can generally
be strictly larger than the union of the set of CIs implied by network
1 and the set of CIs implied by network 2 \cite{del2003qualitative}.\footnote{ The \emph{intersection} of the two sets of CIs given by two Bayesian
network stuctures has also been studied in \cite{matzkevich1993some,del2003qualitative}.
The intersection is the set of CIs that must be satisfied if we know
that the random variables satisfy \emph{at least one} of the network
stuctures. Therefore, our knowledge is weakened by such an intersection
operation. Also, determining whether a CI is implied by such an intersection
is trivial -- simply check whether the CI is in the intersection
\cite{del2003qualitative}. In this paper, we focus on the combination
of knowledge, where we know that the random variables satisfy \emph{both}
network stuctures, corresponding to a gain in knowledge. Determining
whether a CI is implied by such a combination is significantly harder,
as we will see in this paper.}

The \emph{semi-graphoid axioms} by Pearl and Paz \cite{pearl1987graphoids}
is a set of axioms on the implication between conditional independence
relations. One method to combine two Bayesian network structures investigated
in \cite{ron1997combining,del2003qualitative} is to compute the union
of the two aforementioned set of CIs, and then take its closure with
respect to the semi-graphoid axioms. Nevertheless, this set of axioms
is incomplete as shown by Studen\'y \cite{studeny1989multiinformation},
and no finite axiomatization of conditional independence exists \cite{studeny1990conditional}.\footnote{Semi-graphoid axioms can be complete if restricted to certain forms
of conditional independencies, for example, saturated conditional
independencies where each conditional independency must involve all
random variables \cite{malvestuto1992unique,geiger1993logical}. This
fact was utilized in the method proposed in \cite{wong2001constructing}
to combine Markov networks modelling saturated conditional independencies.} Also see \cite{geiger1987non,li2008causal} for some non-axiomatizability
results on Bayesian networks. Furthermore, conditional independence
implication (i.e., given a set of conditional independencies, determine
whether it implies another conditional independency) has recently
been shown to be undecidable by Li \cite{li2023undecidability}, and
also concurrently by K{\"{u}}hne and Yashfe \cite{kuhne2022entropic},
resolving a long-standing open problem.\footnote{The proof in \cite{li2023undecidability} is based on some ideas in
Herrmann's proof of the undecidability of embedded multivalued dependency
\cite{herrmann1995undecidability}. Also refer to \cite{kuhne2019representability,li2022undecidabilityaffine,li2022netcodeundec}
for some related or partial undecidability results on conditional
independence implication.} This eliminates the possibility of the ``first compute the union
of the two sets of CIs, and then algorithmically complete the union''
approach. Even though methods and algorithms more powerful than the
semi-graphoid axioms exist \cite{studeny1989multiinformation,matuvs1995conditional,yeung1997framework,yeung2021machine,li2023automated,boege2024self},
these methods cannot be complete.

In this paper, we show that it is impossible to algorithmically combine
the knowledge of two Bayesian network structures, by proving that
the problem of determining whether two given network structures imply
a given conditional independency  is undecidable, that is, no algorithm
can take two network structures and a CI as inputs, and output whether
the CI is implied by the two network structures. This not only means
that two Bayesian network structures cannot be accurately combined
into one Bayesian network structure, but also means that two Bayesian
network structures cannot be algorithmically combined into \emph{any
data structure} that is capable of answering conditional independency
queries, even if we allow structures that are more general than Bayesian
networks. An interesting consequence of our main result is that we
can explicitly construct two Bayesian network structures, such that
whether they imply a certain CI is unprovable in the ZFC set theory,
assuming ZFC is consistent.\footnote{A Turing machine that does not halt, though the fact that it does
not halt cannot be proved in ZFC, can be constructed \cite{yedidia2016relatively,michel2009busy}.
Since the undecidability proof in this paper is by a reduction from
the undecidability of conditional independence implication \cite{li2023undecidability},
which in turn was shown by a reduction from the uniform word problem
for finite monoids in \cite{gurevich1966problem}, and ultimately
from the halting problem, it is possible to use this sequence of reductions
to construct two Bayesian network structures given the aforementioned
Turing machine, though such networks would be enormous.}

We emphasize that our result is not a direct corollary of the undecidability
of conditional independence implication \cite{li2023undecidability,kuhne2022entropic}
(though we utilize some results in \cite{li2022undecidabilityaffine,li2023undecidability}
in our proof). The result in \cite{li2023undecidability} only implies
that it is undecidable to determine whether a given CI is implied
by a collection of network structures, where the number of network
structures is \emph{unbounded}. This is because the number of CIs
that appear in the formula in \cite{li2023undecidability} is unbounded,
and require an unbounded number of Bayesian networks to impose. This
makes such a hardness result detached from practical scenarios where
the number of Bayesian networks to be combined is often small. The
difficulty in the proof in this paper lies in limiting the number
of Bayesian networks to only two. To this end, we require some novel
constructions for imposing functional dependencies on sufficient statistics
that are not present in \cite{li2022undecidabilityaffine,li2023undecidability}.

We review some other hardness results about probabilistic inference
with Bayesian networks. Cooper \cite{cooper1990computational} showed
that the computation of the marginal distribution of a random variable
in a Bayesian network is NP-hard. Dagum and Luby \cite{dagum1993approximating}
showed that, assuming $\mathrm{P}\neq\mathrm{NP}$, there is no polynomial-time
algorithm for approximating the conditional distribution between two
random variables in a Bayesian network. Also see \cite{roth1996hardness}
for another hardness result. Note that these hardness results are
quantititave -- they are about the computation of probabilities,
and hence concern not only the Bayesian network structure (the directed
acyclic graph), but also the parameters (the probabilities). In comparison,
our hardness result is purely qualitative -- it is about the Bayesian
network structure alone. 

In sum, for a single Bayesian network, qualitative questions about
the structure are easy to answer (linear-time solvable using $d$-separation
\cite{pearl1988probabilistic,verma1990causal,geiger1990logic,geiger1990d}),
and quantititave questions about the probabilities are hard (no polynomial-time
algorithm exists assuming $\mathrm{P}\neq\mathrm{NP}$ \cite{cooper1990computational,dagum1993approximating,roth1996hardness});
whereas for two Bayesian networks, even qualitative questions about
the structure are extremely hard (undecidable, not even exponential-time
algorithm exists, regardless of whether $\mathrm{P}=\mathrm{NP}$
or not) as shown in this paper.

\medskip{}

\section{Preliminaries }

All random variables in this paper are assumed to be discrete with
finite support. For a random variable $X$, write its probability
mass function as $p_{X}(x)$. For a collection of random variables
$X_{1},\ldots,X_{n}$ and $\mathcal{A}\subseteq\{1,\ldots,n\}$, we
write $X_{\mathcal{A}}:=(X_{i})_{i\in\mathcal{A}}$ for the joint
random variable of $X_{i}$ where $i\in\mathcal{A}$. The \emph{support}
of a random variable $X$ is denoted as $\mathcal{X}:=\{x:\,p_{X}(x)>0\}$.
The \emph{cardinality} of $X$ is $|\mathcal{X}|$. 

We say that $X$ and $Y$ are conditionally independent given $Z$,
written as $X\perp\!\!\!\perp Y|Z$, if  $p_{X,Y,Z}(x,y,z)p_{Z}(z)=p_{X,Z}(x,z)p_{Y,Z}(y,z)$
for all $x,y,z$. We say that $X_{1},\ldots,X_{n}$ are conditionally
independent given $Z$, written as $X_{1}\perp\!\!\!\perp\cdots\perp\!\!\!\perp X_{n}|Z$,
if $X_{i}\perp\!\!\!\perp X_{\{1,\ldots,n\}\backslash\{i\}}|Z$ for
every $i=1,\ldots,n$. We say that $X_{1}\to X_{2}\to\cdots\to X_{n}$
forms a Markov chain if $X_{i+1}\perp\!\!\!\perp X_{\{1,\ldots,i-1\}}|X_{i}$
for $i=2,\ldots,n-1$.

We say that $X$ \emph{functionally determines} $Y$ \cite{armstrong1974dependency},
written as $X\stackrel{\iota}{\ge}Y$, if $Y=f(X)$ for some function
$f$, or equivalently, $Y\perp\!\!\!\perp Y|X$. We say that $X$
and $Y$ are \emph{informationally equivalent}, written as $X\stackrel{\iota}{=}Y$,
if $X\stackrel{\iota}{\ge}Y$ and $Y\stackrel{\iota}{\ge}X$.\footnote{These notations are adopted from \cite{li2023undecidability,li2021first}.
We avoid the notation $X\to Y$ for functional dependency in \cite{armstrong1974dependency}
since it can be confused with edges in Bayesian networks and Markov
chains, which do not represent functional dependency.}

Given a joint probability mass function $p_{\mathbf{X}}$ of a collection
of random variables $\mathbf{X}=(X_{1},\ldots,X_{n})$, its set of
conditional independencies (CIs) is the set of tuples $(\mathcal{A},\mathcal{C},\mathcal{B})$
with disjoint sets $\mathcal{A},\mathcal{B},\mathcal{C}\subseteq\{1,\ldots,n\}$
where $X_{\mathcal{A}}\perp\!\!\!\perp X_{\mathcal{B}}|X_{\mathcal{C}}$.
It is written as \cite{koller2009probabilistic}
\begin{align*}
\mathcal{I}(p_{\mathbf{X}}) & :=\big\{(\mathcal{A},\mathcal{C},\mathcal{B})\;\text{disjoint}:\,X_{\mathcal{A}}\perp\!\!\!\perp X_{\mathcal{B}}|X_{\mathcal{C}}\big\}.
\end{align*}

A \emph{Bayesian network structure}  is a directed acyclic graph
$G$ \cite{pearl1988probabilistic}. The set of \emph{local conditional
independencies} of $G=(\mathcal{V},\mathcal{E})$ is \cite{pearl1988probabilistic,koller2009probabilistic}
\[
\mathcal{I}_{\ell}(G):=\big\{(\{i\},\,\mathrm{pa}_{G}(i),\,\mathrm{ndes}_{G}(i)):\,i\in\mathcal{V}\big\},
\]
where $\mathrm{pa}_{G}(i)$ is the set of parents (in-neighbors) of
node $i$ in the graph $G$, and $\mathrm{ndes}_{G}(i)$ is the set
of nodes that are not a descendant or parent of node $i$. Given a
joint probability mass function $p_{\mathbf{X}}$, we say that $p_{\mathbf{X}}$
\emph{satisfies} the Bayesian network structure $G$ with set of nodes
$\{1,\ldots,n\}$, or $G$ is an \emph{independency map (I-map)} for
$p_{\mathbf{X}}$ \cite{pearl1988probabilistic,koller2009probabilistic},
if
\[
\mathcal{I}_{\ell}(G)\subseteq\mathcal{I}(p_{\mathbf{X}}),
\]
or equivalently, $X_{i}\perp\!\!\!\perp X_{\mathrm{ndes}_{G}(i)}|X_{\mathrm{pa}_{G}(i)}$
holds for all node $i$.\footnote{In the literature, a Bayesian network often refers to a pair $(G,p_{\mathbf{X}})$
where $\mathcal{I}_{\ell}(G)\subseteq\mathcal{I}(p_{\mathbf{X}})$.
The directed acyclic graph $G$ is called the \emph{structure} (hence
we call $G$ a Bayesian network structure), and the distribution $p_{\mathbf{X}}$
(or its factorized form \eqref{eq:px_factor}) is called the \emph{parameters}
\cite{pearl1988probabilistic,koller2009probabilistic,theodoridis2020machine}.} We have $\mathcal{I}_{\ell}(G)\subseteq\mathcal{I}(p_{\mathbf{X}})$
if and only if $p_{\mathbf{X}}$ can be factorized as \cite{pearl1988probabilistic,theodoridis2020machine}
\begin{equation}
p_{\mathbf{X}}(x_{1},\ldots,x_{n})=\prod_{i=1}^{n}p_{X_{i}|X_{\mathrm{pa}_{G}(i)}}(x_{i}\,|\,(x_{j})_{j\in\mathrm{pa}_{G}(i)}).\label{eq:px_factor}
\end{equation}

Given a set of CIs $\mathcal{S}\subseteq(2^{\{1,\ldots,n\}})^{3}$,
we say that $\mathcal{S}$ \emph{implies} the CI $(\mathcal{A},\mathcal{C},\mathcal{B})$
(where $\mathcal{A},\mathcal{B},\mathcal{C}\subseteq\{1,\ldots,n\}$
are disjoint) if for every $p_{\mathbf{X}}$ satisfying $X_{\mathcal{A}'}\perp\!\!\!\perp X_{\mathcal{B}'}|X_{\mathcal{C}'}$
for all $(\mathcal{A}',\mathcal{C}',\mathcal{B}')\in\mathcal{S}$,
we must have $X_{\mathcal{A}}\perp\!\!\!\perp X_{\mathcal{B}}|X_{\mathcal{C}}$
\cite{geiger1993logical}. The set of all CIs implied by $\mathcal{S}$
is given by the following ``closure'' operation:
\[
\mathrm{Cl}(\mathcal{S}):=\bigcap_{p_{\mathbf{X}}:\,\mathcal{S}\subseteq\mathcal{I}(p_{\mathbf{X}})}\mathcal{I}(p_{\mathbf{X}}).
\]
where ``$p_{\mathbf{X}}:\,\mathcal{S}\subseteq\mathcal{I}(p_{\mathbf{X}})$''
ranges over all joint probability mass functions of $X_{1},\ldots,X_{n}$
where $\mathcal{S}\subseteq\mathcal{I}(p_{\mathbf{X}})$. In particular,
we say that the Bayesian network structure $G$ \emph{implies} the
CI $(\mathcal{A},\mathcal{C},\mathcal{B})$ , if for every joint probability
mass function $p_{\mathbf{X}}$ satisfying $G$, we must have $X_{\mathcal{A}}\perp\!\!\!\perp X_{\mathcal{B}}|X_{\mathcal{C}}$.
The set of all CIs implied by $G$ is denoted as
\begin{align*}
\mathcal{I}(G) & :=\mathrm{Cl}(\mathcal{I}_{\ell}(G))\\
 & =\Big\{(\mathcal{A},\mathcal{C},\mathcal{B})\;\text{disjoint}:\,\forall p_{\mathbf{X}}.\\
 & \qquad\qquad(\mathcal{I}_{\ell}(G)\subseteq\mathcal{I}(p_{\mathbf{X}}))\to(X_{\mathcal{A}}\perp\!\!\!\perp X_{\mathcal{B}}|X_{\mathcal{C}})\Big\},
\end{align*}
where ``$p_{\mathbf{X}}$'' quantifies over all joint probability
mass functions of $X_{1},\ldots,X_{n}$. It is straightforward to
generalize this definition to a collection of Bayesian network structures
$G_{1},\ldots,G_{k}$. The set of all CIs implied by the collection
$G_{1},\ldots,G_{k}$ is denoted as
\begin{align*}
\mathcal{I}(G_{1},\ldots,G_{k}) & :=\mathrm{Cl}\big(\mathcal{I}_{\ell}(G_{1})\cup\cdots\cup\mathcal{I}_{\ell}(G_{k})\big)\\
 & =\Big\{(\mathcal{A},\mathcal{C},\mathcal{B})\;\text{disjoint}:\,\forall p_{\mathbf{X}}.\\
 & \qquad\;\big(\forall i.\,\mathcal{I}_{\ell}(G_{i})\subseteq\mathcal{I}(p_{\mathbf{X}})\big)\to(X_{\mathcal{A}}\perp\!\!\!\perp X_{\mathcal{B}}|X_{\mathcal{C}})\Big\}.
\end{align*}
In other words, $\mathcal{I}(G_{1},\ldots,G_{k})$ is the set of CIs
that are guaranteed to hold if we know that $p_{\mathbf{X}}$ satisfies
every Bayesian network structure in $G_{1},\ldots,G_{k}$.

\medskip{}

\section{Undecidability Results}

For a single Bayesian network structure $G$, the celebrated $d$\emph{-separation}
method \cite{pearl1988probabilistic,verma1990causal,geiger1990logic,geiger1990d,meek1995strong}
is a sound and complete characterization of all CIs implied by a Bayesian
network structure $G$:
\[
\mathcal{I}(G)=\big\{(\mathcal{A},\mathcal{C},\mathcal{B}):\,\mathcal{A},\mathcal{B}\;\text{are d-separated given}\;\mathcal{C}\;\text{in}\;G\big\},
\]
and hence one Bayesian network structure has decidable CIs. Moreover,
the $d$-separation algorithm is efficient, with a $O(|E|)$ running
time complexity, where $|E|$ is the number of edges \cite{geiger1990d}.

\medskip{}

\begin{thm}
[\cite{verma1990causal,geiger1990logic,geiger1990d}]\label{thm:one_bayes}The
following problem is decidable: Given $n\in\mathbb{Z}_{+}$, a directed
acyclic graph $G$ with set of nodes $\{1,\ldots,n\}$, and disjoint
$\mathcal{A},\mathcal{B},\mathcal{C}\subseteq\{1,\ldots,n\}$, determine
whether $(\mathcal{A},\mathcal{C},\mathcal{B})\in\mathcal{I}(G)$,
i.e., whether the Bayesian network structure $G$ implies the conditional
independency $X_{\mathcal{A}}\perp\!\!\!\perp X_{\mathcal{B}}|X_{\mathcal{C}}$.
\end{thm}
\medskip{}

Next, we study the set of CIs $\mathcal{I}(G_{1},G_{2})$ implied
by two Bayesian network structures $G_{1},G_{2}$. Generally $\mathcal{I}(G_{1},G_{2})\neq\mathcal{I}(G_{1})\cup\mathcal{I}(G_{2})$
\cite{del2003qualitative}. We find out that $\mathcal{I}(G_{1},G_{2})$
is impossible to compute by an algorithm. We now present the main
result in this paper, which shows the undecidability of CIs implied
by two Bayesian network structures. The proof will be presented in
Sections \ref{sec:sufficient} to \ref{sec:construction}.

\medskip{}

\begin{thm}
\label{thm:two_bayes}The following problem is undecidable: Given
$n\in\mathbb{Z}_{+}$, two directed acyclic graphs $G_{1},G_{2}$
with set of nodes $\{1,\ldots,n\}$, and disjoint $\mathcal{A},\mathcal{B},\mathcal{C}\subseteq\{1,\ldots,n\}$,
determine whether $(\mathcal{A},\mathcal{C},\mathcal{B})\in\mathcal{I}(G_{1},G_{2})$,
i.e., whether the Bayesian network structures $G_{1}$ and $G_{2}$
imply the conditional independency $X_{\mathcal{A}}\perp\!\!\!\perp X_{\mathcal{B}}|X_{\mathcal{C}}$.
This problem is still undecidable if we restrict $\mathcal{A},\mathcal{B},\mathcal{C}$
to be singleton sets.\footnote{Although we focus on discrete random variables with finite cardinalities
in this paper, the problem in Theorem \ref{thm:two_bayes} continues
to be undecidable if the discrete random variables are with at most
countable cardinalities. This is because Lemma \ref{lem:smaller_support}
discussed later can ensure that all random variables of interest in
the construction are of finite cardinalities.}
\end{thm}
\medskip{}

Theorem \ref{thm:two_bayes} is in stark contract to Theorem \ref{thm:one_bayes}.
For one Bayesian network structure, CIs can be determined efficiently
in $O(|E|)$ time. For two Bayesian network structures, CIs cannot
be determined by any algorithm, not even exponential time algorithms.
There is no analogue of the $d$-separation algorithm for two Bayesian
networks.

We can also study the implication between Bayesian network structures.
The implication problem with one antecedent in the form ``whether
$G_{1}$ implies $G_{0}$'' is known as the \emph{inclusion problem}
\cite{meek1997graphical,kocka2013characterizing}, which is clearly
decidable due to the $d$-separation algorithm \cite{geiger1990d}.
As an immediate corollary of Theorem \ref{thm:two_bayes}, the Bayesian
network structure implication problem with two antecedents in the
form ``whether $G_{1}$ and $G_{2}$ imply $G_{0}$'' is undecidable.

\medskip{}

\begin{cor}
\label{cor:bayes_impl}Consider the following problem for a fixed
$k$: Given directed acyclic graphs $G_{0},G_{1},\ldots,G_{k}$ with
set of nodes $\{1,\ldots,n\}$, determine whether the Bayesian network
structures $G_{1},\ldots,G_{k}$ imply the Bayesian network structure
$G_{0}$, i.e., whether 
\[
\mathcal{I}(G_{0})\subseteq\mathcal{I}(G_{1},\ldots,G_{k}).
\]
This problem is decidable when $k=1$, and undecidable when $k\ge2$.
\end{cor}
\medskip{}

\begin{IEEEproof}
When $k=1$, we can check whether $G_{1}$ imply $G_{0}$ by computing
$\mathcal{I}(G_{0}),\mathcal{I}(G_{1})$ using the $d$-separation
algorithm \cite{geiger1990d}. When $k=2$, this problem is undecidable
via a reduction from Theorem \ref{thm:two_bayes} by taking $G_{0}$
to be the Bayesian network that imposes only the CI $X_{\mathcal{A}}\perp\!\!\!\perp X_{\mathcal{B}}|X_{\mathcal{C}}$,
for example, a network with edges within each of the sets $\mathcal{A},\mathcal{B},\mathcal{C},\{1,\ldots,n\}\backslash(\mathcal{A}\cup\mathcal{B}\cup\mathcal{C})$,
and with edges $i\to j$ for $(i,j)\in\mathcal{C}\times\mathcal{A}$,
$(i,j)\in\mathcal{C}\times\mathcal{B}$, and $(i,j)\in(\mathcal{A}\cup\mathcal{B}\cup\mathcal{C})\times\{1,\ldots,n\}\backslash(\mathcal{A}\cup\mathcal{B}\cup\mathcal{C})$.
\end{IEEEproof}
\medskip{}

\section{Sufficient Statistics\label{sec:sufficient}}

Before we present the proof of the main result, we review the basic
notion of sufficient statistics \cite{dawid1979conditional}.
\begin{defn}
[Sufficient statistic \cite{dawid1979conditional}] Given random
variables $X,U,T$, we say that $T$ is a \emph{sufficient statistic}
of $X$ for $U$ if $X\stackrel{\iota}{\ge}T$ and $U\perp\!\!\!\perp X|T$,
or equivalently, the Markov chain $U\to T\to X\to T$ holds. We say
that $T^{*}$ is the \emph{minimal sufficient statistic} of $X$ for
$U$ if it is a sufficient statistic of $X$ for $U$, and for every
sufficient statistic $T$ of $X$ for $U$, we have $T\stackrel{\iota}{\ge}T^{*}$.
We denote the minimal sufficient statistic of $X$ for $U$ as
\[
T^{*}=\mathrm{S}(X;U).
\]
\end{defn}
\medskip{}

In the context of Bayesian statistics, $U$ would be a parameter to
be estimated, and $X$ would be an observed sample. $T$ being a sufficient
statistic means that all information about $U$ that can be obtained
from $X$ is contained in $T$.

Note that if $X,U$ are discrete, the minimal sufficient statistic
is given by the random vector $T^{*}=(p_{U|X}(u|X))_{u\in\mathcal{U}}$
with entries indexed by $u\in\mathcal{U}$ (where $\mathcal{U}$ is
the support of $U$). Minimal sufficient statistic is unique, in the
sense that any two minimal sufficient statistics are informationally
equivalent. We then show a basic property of minimal sufficient statistics,
showing that if we are to find the minimal sufficient statistic of
$X$ for $U$, then including an additional piece of information $Y$
that is obtained by processing $X$ locally would not change the minimal
sufficient statistic.

\medskip{}

\begin{lem}
\label{lem:sufficient_add}If $U\perp\!\!\!\perp Y|X$, then $\mathrm{S}(X;U)\stackrel{\iota}{=}\mathrm{S}((X,Y);U)$.
\end{lem}
\begin{IEEEproof}
Let $T=\mathrm{S}((X,Y);U)=(p_{U|X,Y}(u|X,Y))_{u\in\mathcal{U}}$.
Since $U\perp\!\!\!\perp Y|X$, we have $p_{U|X,Y}(u|X,Y)=p_{U|X}(u|X)$,
and hence $T=(p_{U|X}(u|X))_{u\in\mathcal{U}}\stackrel{\iota}{=}\mathrm{S}(X;U)$. 
\end{IEEEproof}
\medskip{}

We then show that we can insert sufficient statistics into Markov
chains.

\medskip{}

\begin{lem}
\label{lem:sufficient_insert}If $V\to U\to X\to Y$ forms a Markov
chain, and $T$ is a sufficient statistic of $X$ for $U$, then $V\to U\to T\to X\to Y$
forms a Markov chain.
\end{lem}
\begin{IEEEproof}
Let $T=f(X)$ (such $f$ exists since $X\stackrel{\iota}{\ge}T$).
We have $p_{V,U,X,Y}(v,u,x,y)=p_{V,U}(v,u)p_{X|U}(x|u)p_{Y|X}(y|x)$,
and $p_{X|U}(x|u)=p_{T|U}(t|u)p_{X|T}(x|t)$. Hence, for $v,u,t,x,y$
with $p_{V,U,T,X,Y}(v,u,t,x,y)>0$, we have $t=f(x)$, and
\begin{align*}
p_{V,U,T,X,Y}(v,u,t,x,y) & =p_{V,U,X,Y}(v,u,x,y)\\
 & =p_{V,U}(v,u)p_{X|U}(x|u)p_{Y|X}(y|x)\\
 & =p_{V,U}(v,u)p_{T|U}(t|u)p_{X|T}(x|t)p_{Y|X}(y|x),
\end{align*}
which shows the desired result.
\end{IEEEproof}
\medskip{}

The purpose of discussing the concept of sufficient statistics is
that, in the Bayesian networks constructed in Section \ref{sec:construction},
we will impose conditions not directly on the variables, but on their
sufficient statistics. It is easier to impose functional dependencies
on sufficient statistics compared to the variables in the network.

\medskip{}

\section{Majorization\label{sec:majorization}}

We also require some tools about the notion of majorization \cite{marshall1979inequalities}.
Given two probability mass functions $p,q$ over the supports $\mathcal{X},\mathcal{Y}$,
respectively, we say that $p$ \emph{majorizes} $q$, written as $p\succeq q$,
if
\[
\sum_{i=1}^{k}p^{\downarrow}(i)\ge\sum_{i=1}^{k}q^{\downarrow}(i),
\]
for every $k\ge1$, where $p^{\downarrow}:\mathbb{Z}_{>0}\to[0,1]$,
$p^{\downarrow}(i)$ is the $i$-th largest entry of $p$ (if there
are fewer than $i$ entries, take $p^{\downarrow}(i)=0$). Majorization
is a partial order over probability mass functions modulo relabeling
\cite{marshall1979inequalities} ($p\succeq q\succeq p$ if and only
if $p^{\downarrow}=q^{\downarrow}$). Also note that if $p\succeq q$,
then the cardinalities satisfy $|\mathcal{X}|\le|\mathcal{Y}|$. 

We prove the following useful result.

\medskip{}

\begin{lem}
\label{lem:smaller_support}Consider random variables $X,Y,Z$ with
probability mass functions $p_{X},p_{Y},p_{Z}$, respectively. We
have the following:
\begin{itemize}
\item If $X\perp\!\!\!\perp Z$ and $X\stackrel{\iota}{\le}(Y,Z)$, then
$p_{X}\succeq p_{Y}$. 
\item If, in addition to the above, we have $X\perp\!\!\!\perp Y$ and $p_{Y}\succeq p_{X}$,
then $X$ and $Y$ are uniformly distributed with the same cardinality,
$Y\perp\!\!\!\perp Z$ and $Y\stackrel{\iota}{\le}(X,Z)$.
\end{itemize}
\end{lem}
\smallskip{}

We remark that although we focus on discrete random variables with
finite cardinalities in this paper, Lemma \ref{lem:smaller_support}
continues to hold for discrete random variables with at most countable
cardinalities. Note that a uniformly distributed discrete random variable
must have a finite cardinality.

\smallskip{}

\begin{IEEEproof}
Since $X\stackrel{\iota}{\le}(Y,Z)$, we can let $X=f_{Z}(Y)$ for
some functions $f_{z}:\mathcal{Y}\to\mathcal{X}$ for $z\in\mathcal{Z}$.
For any subset $\mathcal{A}\subseteq\mathcal{Y}$ with $|\mathcal{A}|\le k$,
\begin{align}
p_{Y}(\mathcal{A}) & =\sum_{z\in\mathcal{Z}}p_{Z}(z)p_{Y|Z=z}(\mathcal{A})\nonumber \\
 & \stackrel{(a)}{\le}\sum_{z\in\mathcal{Z}}p_{Z}(z)p_{X}(f_{z}(\mathcal{A}))\nonumber \\
 & \le\max_{\mathcal{B}\subseteq\mathcal{X}:\,|\mathcal{B}|\le k}p_{X}(\mathcal{B}),\label{eq:maj_ineq}
\end{align}
where (a) is because $Y\in\mathcal{A}$ implies $X=f_{Z}(Y)\in f_{z}(\mathcal{A})$,
and hence $p_{Y|Z=z}(\mathcal{A})\le p_{X|Z=z}(f_{z}(\mathcal{A}))=p_{X}(f_{z}(\mathcal{A}))$.
Therefore $\max_{\mathcal{A}\subseteq\mathcal{Y}:\,|\mathcal{A}|\le k}p_{Y}(\mathcal{A})\le\max_{\mathcal{B}\subseteq\mathcal{X}:\,|\mathcal{B}|\le k}p_{X}(\mathcal{B})$,
and $p_{X}\succeq p_{Y}$.

Assume we also have $X\perp\!\!\!\perp Y$ and $p_{Y}\succeq p_{X}$.
Then $p_{Y}^{\downarrow}=p_{X}^{\downarrow}$. Without loss of generality,
assume $X,Y\in\mathbb{Z}_{>0}$, $p_{X}(x)$ is sorted in descending
order, and $p_{Y}(y)$ is also sorted in descending order. Equality
must hold in the inequalities in \eqref{eq:maj_ineq} when $\mathcal{A}=\mathcal{B}=\{1,\ldots,k\}$.
Therefore,
\begin{align*}
\sum_{z\in\mathcal{Z}}p_{Z}(z)p_{Y|Z=z}(\{1,\ldots,k\}) & =\sum_{z\in\mathcal{Z}}p_{Z}(z)p_{X}(f_{z}(\{1,\ldots,k\}))\\
 & =\sum_{z\in\mathcal{Z}}p_{Z}(z)p_{X}(\{1,\ldots,k\}).
\end{align*}
This implies that for each $z\in\mathcal{Z}$,
\[
p_{Y|Z=z}(\{1,\ldots,k\})=p_{X}(f_{z}(\{1,\ldots,k\}))=p_{X}(\{1,\ldots,k\}).
\]
Taking the difference between the above expressions substituted with
$k$ and $k-1$, we have
\begin{align}
p_{Y|Z=z}(k) & =p_{X}(f_{z}(\{1,\ldots,k\})\backslash f_{z}(\{1,\ldots,k-1\}))\nonumber \\
 & =p_{X}(k).\label{eq:pyzk}
\end{align}
Hence, $p_{Y|Z=z}(k)=p_{X}(k)$, implying that $Y\perp\!\!\!\perp Z$
and the support of $Y$ is $\mathcal{Y}=\mathcal{X}$. Also, whenever
$p_{X}(k)>0$, we have $f_{z}(\{1,\ldots,k\})\backslash f_{z}(\{1,\ldots,k-1\})\neq\emptyset$.
Therefore, $f_{z}(k)$ is injective  over $k\in\mathcal{X}$, and
$Y\stackrel{\iota}{\le}(X,Z)$. By \eqref{eq:pyzk}, for $k\in\mathcal{X}$,
we have $p_{X}(f_{z}(k))=p_{X}(k)$. Since $X\perp\!\!\!\perp Y$,
$X\perp\!\!\!\perp Z$ and $Y\perp\!\!\!\perp Z$, we have, for every
$x,y\in\mathcal{X}$,
\begin{align*}
0<p_{X}(x) & =p_{X|Y=y}(x)\\
 & =\sum_{z\in\mathcal{Z}}p_{Z|Y=y}(z)p_{X|Y=y,Z=z}(x)\\
 & =\sum_{z\in\mathcal{Z}:\,f_{z}(y)=x}p_{Z}(z).
\end{align*}
Hence, for every $x,y\in\mathcal{X}$, there exists $z\in\mathcal{Z}$
such that $f_{z}(y)=x$, which gives $p_{X}(x)=p_{X|Z=z}(x)=p_{X|Z=z}(f_{z}(y))=p_{Y|Z=z}(y)=p_{Y}(y)$
since $f_{z}$ is injective. Therefore, $X$ and $Y$ are uniformly
distributed with the same cardinality.
\end{IEEEproof}
\medskip{}

\section{Special Forms of the Conditional Independence Implication Problem\label{sec:ci_implication}}

Our strategy for proving Theorem \ref{thm:two_bayes} is to show a
reduction from the following special form of the conditional independence
implication problem, which has recently been shown to be undecidable
\cite{li2023undecidability}. It can be proved by noting that the
construction of the undecidable family of conditional independence
implication problems in \cite[Theorem 1]{li2023undecidability} can
be written in the form below. The arguments are briefly described
in Appendix \ref{subsec:pf_three_undec}.

\medskip{}

\begin{thm}
[\cite{li2023undecidability}]\label{thm:three_undec}The following
problem is undecidable: Given $n\ge3$, $\mathcal{A}_{j}\subseteq\{1,\ldots,n\}$
and $b_{j}\in\{1,\ldots,n\}\backslash\mathcal{A}_{j}$ for $j=0,\ldots,k$
where $\mathcal{A}_{0}$ is a singleton set, and $c_{j,0},c_{j,1}\in\{1,\ldots,n\}$,
$c_{j,0}\neq c_{j,1}$ for $j=1,\ldots,k'$, determine whether the
following implication holds: if:
\begin{itemize}
\item each of $V_{1},\ldots,V_{n}$ is uniformly distributed with the same
cardinality,
\item $V_{1}\perp\!\!\!\perp V_{2}\perp\!\!\!\perp V_{3}$,
\item $V_{i}\stackrel{\iota}{\le}(V_{1},V_{2},V_{3})$ for all $i$, 
\item $V_{b_{j}}\stackrel{\iota}{\le}V_{\mathcal{A}_{j}}$ for $j=1,\ldots,k$,
and
\item $V_{c_{j,0}}\perp\!\!\!\perp V_{c_{j,1}}$ for $j=1,\ldots,k'$,
\end{itemize}
then we must have $V_{b_{0}}\stackrel{\iota}{\le}V_{\mathcal{A}_{0}}$.
\end{thm}
\medskip{}

We want to construct Bayesian networks that impose  the antecedents
in Theorem \ref{thm:three_undec}. The condition $V_{1}\perp\!\!\!\perp V_{2}\perp\!\!\!\perp V_{3}$
is easy to impose since we can simply make a Bayesian network where
the three nodes $V_{1},V_{2},V_{3}$ all have no parents. There are
three remaining difficulties. First, the functional dependencies $V_{i}\stackrel{\iota}{\le}(V_{1},V_{2},V_{3})$
and $V_{b_{j}}\stackrel{\iota}{\le}V_{\mathcal{A}_{j}}$ cannot be
imposed by Bayesian networks, since a collection of independent random
variables satisfies any Bayesian network, but does not satisfy any
functional dependency between the variables. This problem can be solved
by considering sufficient statistics described in Section \ref{sec:sufficient}.
Second, we have to impose  that each of $V_{1},\ldots,V_{n}$ is uniformly
distributed with the same cardinality. Third, the conditions $V_{c_{j,0}}\perp\!\!\!\perp V_{c_{j,1}}$
for $j=1,\ldots,k'$ would require $k'$ Bayesian networks to impose.
To this end, we utilize the following lemma.

\medskip{}

\begin{lem}
\label{lem:iid_extend}Let $V\sim\mathrm{Unif}(\{1,\ldots\ell^{n}\})$
for some $\ell,n\ge1$. Assume the random variables $X_{1},\ldots,X_{k}$
($k\le n$) with cardinalities $|\mathcal{X}_{i}|\le\ell$ for $i=1,\ldots,k$,
satisfies that $(X_{1},\ldots,X_{k})\stackrel{\iota}{\le}V$. Then
the following statements are equivalent:
\begin{enumerate}
\item $X_{1},\ldots,X_{k}$ are mutually independent, each being uniformly
distributed with cardinality $\ell$.
\item There exists random variables $Y_{1},\ldots,Y_{n-k}$ satisfying $|\mathcal{Y}_{i}|\le\ell$
for $i=1,\ldots,n-k$ and $((X_{i})_{i=1}^{k},(Y_{i})_{i=1}^{n-k})\stackrel{\iota}{=}V$.
\item There exists random variables $Y_{1},\ldots,Y_{n-k}$ each being uniformly
distributed with cardinality $\ell$, satisfying $((X_{i})_{i=1}^{k},(Y_{i})_{i=1}^{n-k})\stackrel{\iota}{=}V$.
\end{enumerate}
\end{lem}
\begin{IEEEproof}
Statement 3 $\Rightarrow$ Statement 2 is obvious. For Statement 2
$\Rightarrow$ Statement 1, assume $Y_{1},\ldots,Y_{n-k}$ with $|\mathcal{Y}_{i}|\le\ell$
for $i=1,\ldots,n-k$ and $((X_{i})_{i},(Y_{i})_{i})\stackrel{\iota}{=}V$.
The number of possibilities of $((X_{i})_{i},(Y_{i})_{i})$ is at
most $\ell^{n}$. Since $((X_{i})_{i},(Y_{i})_{i})\stackrel{\iota}{=}V$
and $|\mathcal{V}|=\ell^{n}$, there is a bijection between the values
of $((X_{i})_{i},(Y_{i})_{i})$ and the values of $V$. Hence, all
$\ell^{n}$ possibilities of $((X_{i})_{i},(Y_{i})_{i})$ are equally
likely, and hence $(X_{i})_{i},(Y_{i})_{i}$ are mutually independent,
each being uniform with cardinality $\ell$.

For Statement 1 $\Rightarrow$ Statement 3, consider $X_{1},\ldots,X_{k}$
mutually independent, each being uniformly distributed with cardinality
$\ell$. We construct $Y_{1},\ldots,Y_{n-k}$ as follows. Since $(X_{1},\ldots,X_{k})\stackrel{\iota}{\le}V$,
we can find $f:\{1,\ldots,\ell^{n}\}\to\{1,\ldots,\ell\}^{k}$ such
that $(X_{i})_{i}=f(V)$. Since $(X_{i})_{i}$ is uniform over $\ell^{k}$
choices, the preimage $f^{-1}((x_{i})_{i})$ has size $|f^{-1}((x_{i})_{i})|=\ell^{n-k}$
for each $(x_{i})_{i}\in\prod_{i}\mathcal{X}_{i}$. Let $f^{-1}((x_{i})_{i})=\{g((x_{i})_{i},y):\,y\in\{1,\ldots,\ell^{n-k}\}\}$.
Then $g:(\prod_{i}\mathcal{X}_{i})\times\{1,\ldots,\ell^{n-k}\}\to\{1,\ldots,\ell^{n}\}$
is bijective, and has an inverse $g^{-1}$. Take $Y=(g^{-1}(V))_{2}\in\{1,\ldots,\ell^{n-k}\}$
(the second component of $g^{-1}(V)$). We have $Y\sim\mathrm{Unif}(\{1,\ldots,\ell^{n-k}\})$
and $((X_{i})_{i},Y)\stackrel{\iota}{=}V$. The proof is completed
by breaking $Y$ into $n-k$ independent uniform random variables
$(Y_{1},\ldots,Y_{n-k})\stackrel{\iota}{=}Y$. 
\end{IEEEproof}
\medskip{}

We now show a useful intermediate result, which is the undecidability
of determining whether two mutual independence relations and some
functional dependencies imply another functional dependency. This
is in contrast to the implication problem of functional dependencies,
which is decidable and finitely axiomatizable \cite{armstrong1974dependency}.
This result might be interesting on its own.

\medskip{}

\begin{thm}
\label{thm:seq_undec}The following problem is undecidable: Given
$n\ge1$, $\mathcal{C}_{1},\mathcal{C}_{2}\subseteq\{1,\ldots,n\}$
with $\mathcal{C}_{1}\cap\mathcal{C}_{2}=\emptyset$, $\mathcal{A}_{j}\subseteq\{1,\ldots,n\}$
and $b_{j}\in\{1,\ldots,n\}\backslash\mathcal{A}_{j}$ for $j=0,\ldots,k$
where $\mathcal{A}_{0}$ is a singleton set, determine whether the
following implication holds: if $V_{1},\ldots,V_{n}$ satisfy that
the random variables $(V_{i})_{i\in\mathcal{C}_{1}}$ are mutually
independent, the random variables $(V_{i})_{i\in\mathcal{C}_{2}}$
are mutually independent, and $V_{b_{j}}\stackrel{\iota}{\le}V_{\mathcal{A}_{j}}$
for $j=1,\ldots,k$, then we must have $V_{b_{0}}\stackrel{\iota}{\le}V_{\mathcal{A}_{0}}$.
\end{thm}
\medskip{}

\begin{IEEEproof}
We first show that the conditions $V_{c_{j,0}}\perp\!\!\!\perp V_{c_{j,1}}$
for $j=1,\ldots,k'$ are unnecessary for the problem in Theorem \ref{thm:three_undec},
i.e., we show that the following implication problem without those
conditions is still undecidable: 

\textbf{Implication A:} ``If $V_{1},\ldots,V_{n}$ satisfy that each
of $V_{1},\ldots,V_{n}$ is uniform with the same cardinality, $V_{1}\perp\!\!\!\perp V_{2}\perp\!\!\!\perp V_{3}$,
$V_{i}\stackrel{\iota}{\le}(V_{1},V_{2},V_{3})$ for all $i$, and
$V_{b_{j}}\stackrel{\iota}{\le}V_{\mathcal{A}_{j}}$ for $j=1,\ldots,k$,
then we must have $V_{b_{0}}\stackrel{\iota}{\le}V_{\mathcal{A}_{0}}$.''

We show that Implication A is undecidable by a reduction from Theorem
\ref{thm:three_undec}. Consider an instance of the implication problem
in Theorem \ref{thm:three_undec}. By Lemma \ref{lem:iid_extend},
if each of $V_{1},\ldots,V_{n}$ is uniform with the same cardinality
and $V_{i}\stackrel{\iota}{\le}(V_{1},V_{2},V_{3})$ for all $i$,
then $V_{c_{j,0}}\perp\!\!\!\perp V_{c_{j,1}}$ if and only if there
exists $Y_{j}$ uniform with the same cardinality such that $(V_{c_{j,0}},V_{c_{j,1}}Y_{j})\stackrel{\iota}{=}(V_{1},V_{2},V_{3})$.
Therefore, we can impose the ``$V_{c_{j,0}}\perp\!\!\!\perp V_{c_{j,1}}$''
conditions by introducing new variables $Y_{j}$ and imposing functional
dependencies ``$(V_{c_{j,0}},V_{c_{j,1}}Y_{j})\stackrel{\iota}{=}(V_{1},V_{2},V_{3})$''
instead. The functional dependency conditions in Implication A suffice
to impose the ``$V_{c_{j,0}}\perp\!\!\!\perp V_{c_{j,1}}$'' conditions.

Next, we show the desired undecidability via a reduction from Implication
A. Consider an instance of Implication A given as $\mathcal{A}_{j}\subseteq\{1,\ldots,n\}$
and $b_{j}\in\{1,\ldots,n\}$ for $j=0,\ldots,k$. We will prove that
this instance of Implication A holds if and only if the following
implication (which is in the form in Theorem \ref{thm:seq_undec})
holds:

\textbf{Implication B:} ``If $Q,V_{1},\dots,V_{3n},W_{1},\ldots,W_{3n+1},M_{1},\ldots,M_{3n}$
satisfy these antecedents:
\begin{itemize}
\item $Q\perp\!\!\!\perp V_{1}\perp\!\!\!\perp V_{2}\perp\!\!\!\perp V_{3}$,
\item $M_{1}\perp\!\!\!\perp M_{2}\perp\!\!\!\perp\cdots\perp\!\!\!\perp M_{3n}\perp\!\!\!\perp W_{3n+1}$,
\item $W_{i}\stackrel{\iota}{\le}(W_{i+1},M_{i})$ and $W_{i+1}\stackrel{\iota}{\le}(W_{i},M_{i})$
for $i=1,\ldots,3n$,
\item $W_{1}\stackrel{\iota}{\le}(W_{3n+1},Q)$,
\item $V_{i}\stackrel{\iota}{=}W_{i}$ for $i=1,2,3$,
\item $V_{i}\stackrel{\iota}{\le}(V_{1},V_{2},V_{3})$ for $i=1,\ldots,3n$, 
\item $V_{i}\stackrel{\iota}{\le}(W_{i},Q)$ for $i=1,\ldots,3n$, 
\item $(V_{1},V_{2},V_{3})\stackrel{\iota}{=}(V_{i},V_{i+n},V_{i+2n})$
for $i=1,\ldots,n$,
\item $V_{b_{j}}\stackrel{\iota}{\le}V_{\mathcal{A}_{j}}$ for $j=1,\ldots,k$, 
\end{itemize}
then we must have $V_{b_{0}}\stackrel{\iota}{\le}V_{\mathcal{A}_{0}}$.''

Some of the functional dependencies are illustrated below (note that
this is not a Bayesian network). A directed hyperedge ``$\begin{array}{c}
X\\
Y
\end{array}\!\!\!\rightarrowtail Z$'' means $(X,Y)\stackrel{\iota}{\ge}Z$. Technically the number of
$M_{i}$'s must be a multiple of $3$, though we ignore this for the
illustration.\\
% https://q.uiver.app/#q=WzAsMTIsWzQsNCwiUSJdLFswLDIsIlZfMVxcc3RhY2tyZWx7XFxpb3RhfXs9fVdfMSJdLFsyLDIsIlZfMlxcc3RhY2tyZWx7XFxpb3RhfXs9fVdfMiJdLFs0LDIsIlZfM1xcc3RhY2tyZWx7XFxpb3RhfXs9fVdfMyJdLFs2LDIsIldfNCJdLFsxLDAsIk1fMSJdLFszLDAsIk1fMiJdLFs1LDAsIk1fMyJdLFs2LDQsIlZfNCJdLFs4LDIsIldfNSJdLFs4LDQsIlZfNSJdLFs3LDAsIk1fNCJdLFs1LDEsIiIsMSx7ImN1cnZlIjoxfV0sWzUsMiwiIiwxLHsiY3VydmUiOi0xfV0sWzYsMiwiIiwxLHsiY3VydmUiOjF9XSxbNiwzLCIiLDEseyJjdXJ2ZSI6LTF9XSxbNywzLCIiLDEseyJjdXJ2ZSI6MX1dLFs3LDQsIiIsMSx7ImN1cnZlIjotMX1dLFsxMSw0LCIiLDEseyJjdXJ2ZSI6MX1dLFsxMSw5LCIiLDEseyJjdXJ2ZSI6LTF9XSxbMCwxXSxbNCw4XSxbOSwxMF0sWzIsMTIsIiIsMSx7ImN1cnZlIjoyLCJsZXZlbCI6MSwic3R5bGUiOnsiaGVhZCI6eyJuYW1lIjoibm9uZSJ9fX1dLFsxLDEzLCIiLDEseyJjdXJ2ZSI6LTIsImxldmVsIjoxLCJzdHlsZSI6eyJoZWFkIjp7Im5hbWUiOiJub25lIn19fV0sWzMsMTQsIiIsMSx7ImN1cnZlIjoyLCJsZXZlbCI6MSwic3R5bGUiOnsiaGVhZCI6eyJuYW1lIjoibm9uZSJ9fX1dLFs0LDE2LCIiLDEseyJjdXJ2ZSI6MiwibGV2ZWwiOjEsInN0eWxlIjp7ImhlYWQiOnsibmFtZSI6Im5vbmUifX19XSxbOSwxOCwiIiwxLHsiY3VydmUiOjIsImxldmVsIjoxLCJzdHlsZSI6eyJoZWFkIjp7Im5hbWUiOiJub25lIn19fV0sWzIsMTUsIiIsMSx7ImN1cnZlIjotMiwibGV2ZWwiOjEsInN0eWxlIjp7ImhlYWQiOnsibmFtZSI6Im5vbmUifX19XSxbMywxNywiIiwxLHsiY3VydmUiOi0yLCJsZXZlbCI6MSwic3R5bGUiOnsiaGVhZCI6eyJuYW1lIjoibm9uZSJ9fX1dLFs0LDE5LCIiLDEseyJjdXJ2ZSI6LTIsImxldmVsIjoxLCJzdHlsZSI6eyJoZWFkIjp7Im5hbWUiOiJub25lIn19fV0sWzksMjAsIiIsMSx7ImN1cnZlIjotMiwibGV2ZWwiOjEsInN0eWxlIjp7ImhlYWQiOnsibmFtZSI6Im5vbmUifX19XSxbMCwyMSwiIiwxLHsiY3VydmUiOi0yLCJsZXZlbCI6MSwic3R5bGUiOnsiaGVhZCI6eyJuYW1lIjoibm9uZSJ9fX1dLFswLDIyLCIiLDEseyJjdXJ2ZSI6LTIsImxldmVsIjoxLCJzdHlsZSI6eyJoZWFkIjp7Im5hbWUiOiJub25lIn19fV1d
\[\begin{tikzcd}[column sep=small]
	& {M_1} && {M_2} && {M_3} && {M_4} \\
	\\
	{V_1\stackrel{\iota}{=}W_1} && {V_2\stackrel{\iota}{=}W_2} && {V_3\stackrel{\iota}{=}W_3} && {W_4} && {W_5} \\
	\\
	&&&& Q && {V_4} && {V_5}
	\arrow[""{name=0, anchor=center, inner sep=0}, curve={height=6pt}, from=1-2, to=3-1]
	\arrow[""{name=1, anchor=center, inner sep=0}, curve={height=-6pt}, from=1-2, to=3-3]
	\arrow[""{name=2, anchor=center, inner sep=0}, curve={height=6pt}, from=1-4, to=3-3]
	\arrow[""{name=3, anchor=center, inner sep=0}, curve={height=-6pt}, from=1-4, to=3-5]
	\arrow[""{name=4, anchor=center, inner sep=0}, curve={height=6pt}, from=1-6, to=3-5]
	\arrow[""{name=5, anchor=center, inner sep=0}, curve={height=-6pt}, from=1-6, to=3-7]
	\arrow[""{name=6, anchor=center, inner sep=0}, curve={height=6pt}, from=1-8, to=3-7]
	\arrow[""{name=7, anchor=center, inner sep=0}, curve={height=-6pt}, from=1-8, to=3-9]
	\arrow[""{name=8, anchor=center, inner sep=0}, from=3-7, to=5-7]
	\arrow[""{name=9, anchor=center, inner sep=0}, from=3-9, to=5-9]
	\arrow[""{name=10, anchor=center, inner sep=0}, from=5-5, to=3-1]
	\arrow[curve={height=-12pt}, no head, from=3-1, to=1]
	\arrow[curve={height=12pt}, no head, from=3-3, to=0]
	\arrow[curve={height=-12pt}, no head, from=3-3, to=3]
	\arrow[curve={height=12pt}, no head, from=3-5, to=2]
	\arrow[curve={height=-12pt}, no head, from=3-5, to=5]
	\arrow[curve={height=12pt}, no head, from=3-7, to=4]
	\arrow[curve={height=-12pt}, no head, from=3-7, to=7]
	\arrow[curve={height=-12pt}, no head, from=3-9, to=10]
	\arrow[curve={height=12pt}, no head, from=3-9, to=6]
	\arrow[curve={height=-12pt}, no head, from=5-5, to=8]
	\arrow[curve={height=-12pt}, no head, from=5-5, to=9]
\end{tikzcd}\]Intuitively, the conditions $M_{1}\perp\!\!\!\perp\cdots\perp\!\!\!\perp M_{3n}\perp\!\!\!\perp W_{3n+1}$
and $W_{i}\stackrel{\iota}{\le}(W_{i+1},M_{i})$ gives a Markov chain
$W_{3n+1}\to W_{3n}\to\cdots\to W_{1}$. This together with the dependency
$W_{1}\stackrel{\iota}{\le}(W_{3n+1},Q)$ in the other direction gives
a ``cyclic'' condition that ensures that $W_{1},\ldots,W_{3n}$
are identically distributed. We can then use Lemma \ref{lem:smaller_support}
to argue that the random variables are uniformly distributed.

First prove the ``only if'' direction and assume that Implication
A holds, and let $Q,(V_{i})_{i},(W_{i})_{i},(M_{i})_{i}$ satisfy
the antecedents of Implication B. Note that since $W_{i}\stackrel{\iota}{\le}(W_{i+1},M_{i})$
for $i=1,\ldots,3n$, we have 
\[
W_{i+1}\stackrel{\iota}{\le}(M_{i+1},\ldots,M_{3n},W_{3n+1})\perp\!\!\!\perp M_{i}.
\]
By $W_{i+1}\stackrel{\iota}{\le}(W_{i},M_{i})$, we have $p_{W_{i+1}}\succeq p_{W_{i}}$
for $i=1,\ldots,3n$ by Lemma \ref{lem:smaller_support}. Also since
$W_{1}\stackrel{\iota}{=}V_{1}\perp\!\!\!\perp Q$ and $W_{1}\stackrel{\iota}{\le}(W_{3n+1},Q)$,
we have $p_{W_{1}}\succeq p_{W_{3n+1}}$ by Lemma \ref{lem:smaller_support}.
Therefore, we have $p_{W_{1}}^{\downarrow}=\cdots=p_{W_{3n+1}}^{\downarrow}$.
Since $W_{1}\stackrel{\iota}{=}V_{1}\perp\!\!\!\perp W_{2}\stackrel{\iota}{=}V_{2}$,
$W_{2}\perp\!\!\!\perp M_{1}$ and $W_{2}\stackrel{\iota}{\le}(W_{1},M_{1})$,
$W_{1}$ is uniformly distributed by Lemma \ref{lem:smaller_support}.
Therefore, all of $W_{1},\ldots,W_{3n+1}$ are uniformly distributed
with the same cardinality. Let the cardinality be $\ell$. By Lemma
\ref{lem:smaller_support}, since $V_{i}\stackrel{\iota}{\le}(V_{1},V_{2},V_{3})\perp\!\!\!\perp Q$
and $V_{i}\stackrel{\iota}{\le}(W_{i},Q)$, we have $|\mathcal{V}_{i}|\le\ell$
for $i=1,\ldots,3n$. By Lemma \ref{lem:iid_extend}, since $(V_{1},V_{2},V_{3})\stackrel{\iota}{=}(V_{i},V_{i+n},V_{i+2n})$,
we know that $V_{i}$ is uniform with cardinality $\ell$ for $i=1,\ldots,n$.
Applying Implication A on $V_{1},\ldots,V_{n}$, we have $V_{b_{0}}\stackrel{\iota}{\le}V_{\mathcal{A}_{0}}$.

Then prove the ``if'' direction and assume that Implication B holds,
and let $V_{1},\dots,V_{n}$ satisfy the antecedents in Implication
A. Without loss of generality, assume each of $V_{1},\dots,V_{n}$
has a distribution $\mathrm{Unif}(\{0,\ldots,\ell-1\})$. Since $V_{i}\stackrel{\iota}{\le}(V_{1},V_{2},V_{3})$
and $V_{i}$ is uniform with cardinality $\ell$ for $i=1,\ldots,n$,
we can construct $V_{i+n},V_{i+2n}$ such that $V_{i},V_{i+n},V_{i+2n}\stackrel{iid}{\sim}\mathrm{Unif}(\{0,\ldots,\ell-1\})$,
and $(V_{1},V_{2},V_{3})\stackrel{\iota}{=}(V_{i},V_{i+n},V_{i+2n})$.
Take $Q=(Q_{4},\ldots,Q_{3n+1})$ where $Q_{4},\ldots,Q_{3n+1}\stackrel{iid}{\sim}\mathrm{Unif}(\{0,\ldots,\ell-1\})$,
and $W_{i}=V_{i}$ for $i=1,2,3$, $W_{i}=(V_{i}+Q_{i})\;\mathrm{mod}\;\ell$
for $i=4,\ldots,3n$, $W_{3n+1}=(V_{1}+Q_{3n+1})\;\mathrm{mod}\;\ell$.
We have $V_{i}=(W_{i}-Q_{i})\;\mathrm{mod}\;\ell\stackrel{\iota}{\le}(W_{i},Q)$
for $i=4,\ldots,3n$, and $V_{i}\stackrel{\iota}{\le}(W_{i},Q)$ clearly
holds for $i=1,2,3$. Also, $W_{1}=V_{1}=(W_{3n+1}-Q_{3n+1})\;\mathrm{mod}\;\ell\stackrel{\iota}{\le}(W_{3n+1},Q)$.
Take $M_{i}=(W_{i}+W_{i+1})\;\mathrm{mod}\;\ell$ for $i=1,\ldots,3n$.
We have $W_{i}\stackrel{\iota}{\le}(W_{i+1},M_{i})$ and $W_{i+1}\stackrel{\iota}{\le}(W_{i},M_{i})$.
It is straightforward to check that $W_{1},\ldots,W_{3n+1}\stackrel{iid}{\sim}\mathrm{Unif}(\{0,\ldots,\ell-1\})$,
and hence $M_{1},\ldots,M_{3n},W_{3n+1}\stackrel{iid}{\sim}\mathrm{Unif}(\{0,\ldots,\ell-1\})$.
Hence, the antecedents of Implication B are satisfied, and we have
$V_{b_{0}}\stackrel{\iota}{\le}V_{\mathcal{A}_{0}}$.
\end{IEEEproof}
\medskip{}

\section{Construction of the Bayesian Networks\label{sec:construction}}

We now complete the proof of Theorem \ref{thm:two_bayes} by constructing
the two Bayesian network structures, according to an instance of the
implication problem in Theorem \ref{thm:seq_undec}. For the sake
of convenience of the construction, we impose an additional constraint
on the list of functional dependencies, and consider this problem:\medskip{}

\begin{cor}
\label{cor:undec_eq}The following problem is undecidable: Given $n\ge2$,
$\mathcal{C}_{1},\mathcal{C}_{2}\subseteq\{1,\ldots,n\}$ with $\mathcal{C}_{1}\cap\mathcal{C}_{2}=\emptyset$,
$\mathcal{A}_{j}\subseteq\{1,\ldots,n\}$, $b_{j}\in\{1,\ldots,n\}\backslash\mathcal{A}_{j}$
for $j=0,\ldots,k$ where $\mathcal{A}_{0}$ is a singleton set, and
$b_{0}'\in\{1,\ldots,n\}\backslash(\mathcal{A}_{0}\cup\{b_{0}\})$
such that there are two entries $(\{b_{0}\},b_{0}')$ and $(\{b_{0}'\},b_{0})$
among the list of functional dependencies $(\mathcal{A}_{j},b_{j})_{j=1}^{k}$,
determine whether the following implication holds: if $V_{1},\ldots,V_{n}$
satisfy that the random variables $(V_{i})_{i\in\mathcal{C}_{1}}$
are mutually independent, the random variables $(V_{i})_{i\in\mathcal{C}_{2}}$
are mutually independent, and $V_{b_{j}}\stackrel{\iota}{\le}V_{\mathcal{A}_{j}}$
for $j=1,\ldots,k$, then we must have $V_{b_{0}}\stackrel{\iota}{\le}V_{\mathcal{A}_{0}}$.
\end{cor}
\medskip{}

The purpose of the two entries $(\{b_{0}\},b_{0}')$ and $(\{b_{0}'\},b_{0})$
are to impose $V_{b_{0}}\stackrel{\iota}{\ge}V_{b_{0}'}$ and $V_{b_{0}'}\stackrel{\iota}{\ge}V_{b_{0}}$,
and hence $V_{b_{0}}\stackrel{\iota}{=}V_{b_{0}'}$. Corollary \ref{cor:undec_eq}
clearly follows from Theorem \ref{thm:seq_undec} since we can always
introduce a new random variable $V_{b_{0}'}$ and impose that $V_{b_{0}'}\stackrel{\iota}{=}V_{b_{0}}$,
without affecting the truth value of the original implication problem.
Having two identical copies $V_{b_{0}}\stackrel{\iota}{=}V_{b_{0}'}$
makes it more convenient to check whether $V_{b_{0}}\stackrel{\iota}{\le}V_{\mathcal{A}_{0}}$
in the construction introduced later.

Consider an instance of the implication problem in Corollary \ref{cor:undec_eq}
given as $\mathcal{C}_{1},\mathcal{C}_{2}\subseteq\{1,\ldots,n\}$
with $\mathcal{C}_{1}\cap\mathcal{C}_{2}=\emptyset$, $\mathcal{A}_{j}\subseteq\{1,\ldots,n\}$,
$b_{j}\in\{1,\ldots,n\}\backslash\mathcal{A}_{j}$ for $j=0,\ldots,k$
where $\mathcal{A}_{0}$ is a singleton set, and $b_{0}'\in\{1,\ldots,n\}\backslash(\mathcal{A}_{0}\cup\{b_{0}\})$
such that there are two entries $(\{b_{0}\},b_{0}')$ and $(\{b_{0}'\},b_{0})$
among $(\mathcal{A}_{j},b_{j})_{j=1}^{k}$. The variables in the networks
are $U_{i},Y_{i},Z_{i},X_{i}^{j}$ for $i=1,\ldots,n$ and $j=1,\ldots,k$
(note that the superscript $j$ in $X_{i}^{j}$ is an index to the
two-dimensional array $(X_{i}^{j})_{i,j}$, and is not a power). We
now describe the two networks.
\begin{itemize}
\item Network 1 has edges:
\begin{itemize}
\item $U_{i}\to U_{i'}$ for $i\in\mathcal{C}_{1}$, $i'\in\{1,\ldots,n\}\backslash\mathcal{C}_{1}$,
\item $U_{i}\to U_{i'}$ for $i,i'\in\{1,\ldots,n\}\backslash\mathcal{C}_{1}$
where $i<i'$,
\item $U_{i}\to Y_{i}\to X_{i}^{1}\to\cdots\to X_{i}^{k}\to Z_{i}$ for
$i=1,\ldots,n$.
\end{itemize}
\item Network 2 has edges:
\begin{itemize}
\item $U_{i}\to U_{i'}$ for $i\in\mathcal{C}_{2}$, $i'\in\{1,\ldots,n\}\backslash\mathcal{C}_{2}$,
\item $U_{i}\to U_{i'}$ for $i,i'\in\{1,\ldots,n\}\backslash\mathcal{C}_{2}$
where $i<i'$,
\item $U_{i}\to Z_{i}\to Y_{i}$ for $i=1,\ldots,n$, 
\item $U_{i}\to X_{i}^{j}$ for $j=1,\ldots,k$, $i\in\{1,\ldots,n\}\backslash\{b_{j}\}$,
\item $X_{i}^{j}\to X_{b_{j}}^{j}$ for $j\in1,\ldots,k$, $i\in\mathcal{A}_{j}$.
\end{itemize}
\end{itemize}
\medskip{}

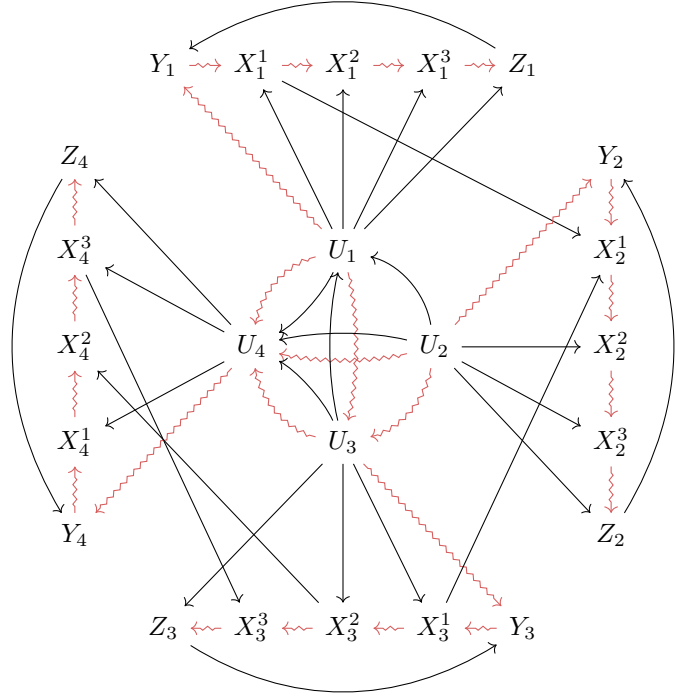
\begin{figure}
\[\begin{tikzcd}[column sep=small]
	& {Y_1} & {X_1^1} & {X_1^2} & {X_1^3} & {Z_1} \\
	{Z_4} &&&&&& {Y_2} \\
	{X_4^3} &&& {U_1} &&& {X_2^1} \\
	{X_4^2} && {U_4} && {U_2} && {X_2^2} \\
	{X_4^1} &&& {U_3} &&& {X_2^3} \\
	{Y_4} &&&&&& {Z_2} \\
	& {Z_3} & {X_3^3} & {X_3^2} & {X_3^1} & {Y_3}
	\arrow[draw={rgb,255:red,214;green,92;blue,92}, squiggly, from=1-2, to=1-3]
	\arrow[draw={rgb,255:red,214;green,92;blue,92}, squiggly, from=1-3, to=1-4]
	\arrow[from=1-3, to=3-7]
	\arrow[color={rgb,255:red,214;green,92;blue,92}, squiggly, from=1-4, to=1-5]
	\arrow[color={rgb,255:red,214;green,92;blue,92}, squiggly, from=1-5, to=1-6]
	\arrow[curve={height=30pt}, from=1-6, to=1-2]
	\arrow[curve={height=30pt}, from=2-1, to=6-1]
	\arrow[draw={rgb,255:red,214;green,92;blue,92}, squiggly, from=2-7, to=3-7]
	\arrow[color={rgb,255:red,214;green,92;blue,92}, squiggly, from=3-1, to=2-1]
	\arrow[from=3-1, to=7-3]
	\arrow[draw={rgb,255:red,214;green,92;blue,92}, squiggly, from=3-4, to=1-2]
	\arrow[from=3-4, to=1-3]
	\arrow[from=3-4, to=1-4]
	\arrow[from=3-4, to=1-5]
	\arrow[from=3-4, to=1-6]
	\arrow[curve={height=-6pt}, from=3-4, to=4-3]
	\arrow[draw={rgb,255:red,214;green,92;blue,92}, curve={height=12pt}, squiggly, from=3-4, to=4-3]
	\arrow[draw={rgb,255:red,214;green,92;blue,92}, curve={height=-6pt}, squiggly, from=3-4, to=5-4]
	\arrow[draw={rgb,255:red,214;green,92;blue,92}, squiggly, from=3-7, to=4-7]
	\arrow[color={rgb,255:red,214;green,92;blue,92}, squiggly, from=4-1, to=3-1]
	\arrow[from=4-3, to=2-1]
	\arrow[from=4-3, to=3-1]
	\arrow[from=4-3, to=5-1]
	\arrow[draw={rgb,255:red,214;green,92;blue,92}, squiggly, from=4-3, to=6-1]
	\arrow[draw={rgb,255:red,214;green,92;blue,92}, squiggly, from=4-5, to=2-7]
	\arrow[curve={height=12pt}, from=4-5, to=3-4]
	\arrow[draw={rgb,255:red,214;green,92;blue,92}, curve={height=-6pt}, squiggly, from=4-5, to=4-3]
	\arrow[curve={height=6pt}, from=4-5, to=4-3]
	\arrow[from=4-5, to=4-7]
	\arrow[draw={rgb,255:red,214;green,92;blue,92}, curve={height=-12pt}, squiggly, from=4-5, to=5-4]
	\arrow[from=4-5, to=5-7]
	\arrow[from=4-5, to=6-7]
	\arrow[color={rgb,255:red,214;green,92;blue,92}, squiggly, from=4-7, to=5-7]
	\arrow[draw={rgb,255:red,214;green,92;blue,92}, squiggly, from=5-1, to=4-1]
	\arrow[curve={height=-6pt}, from=5-4, to=3-4]
	\arrow[draw={rgb,255:red,214;green,92;blue,92}, curve={height=-12pt}, squiggly, from=5-4, to=4-3]
	\arrow[curve={height=6pt}, from=5-4, to=4-3]
	\arrow[from=5-4, to=7-2]
	\arrow[from=5-4, to=7-4]
	\arrow[from=5-4, to=7-5]
	\arrow[draw={rgb,255:red,214;green,92;blue,92}, squiggly, from=5-4, to=7-6]
	\arrow[color={rgb,255:red,214;green,92;blue,92}, squiggly, from=5-7, to=6-7]
	\arrow[draw={rgb,255:red,214;green,92;blue,92}, squiggly, from=6-1, to=5-1]
	\arrow[curve={height=30pt}, from=6-7, to=2-7]
	\arrow[curve={height=30pt}, from=7-2, to=7-6]
	\arrow[color={rgb,255:red,214;green,92;blue,92}, squiggly, from=7-3, to=7-2]
	\arrow[from=7-4, to=4-1]
	\arrow[color={rgb,255:red,214;green,92;blue,92}, squiggly, from=7-4, to=7-3]
	\arrow[from=7-5, to=3-7]
	\arrow[draw={rgb,255:red,214;green,92;blue,92}, squiggly, from=7-5, to=7-4]
	\arrow[draw={rgb,255:red,214;green,92;blue,92}, squiggly, from=7-6, to=7-5]
\end{tikzcd}\]

\caption{\label{fig:networks}An illustration of the two networks when $n=4$,
$k=3$, $\mathcal{C}_{1}=\{1,2\}$, $\mathcal{C}_{1}=\{2,3\}$, $\mathcal{A}_{1}=\{1,3\}$,
$b_{1}=2$, $\mathcal{A}_{2}=\{3\}$, $b_{2}=4$, $\mathcal{A}_{3}=\{4\}$,
$b_{3}=3$, i.e., we impose $V_{1}\perp\!\!\!\perp V_{2}$, $V_{2}\perp\!\!\!\perp V_{3}$,
$V_{2}\stackrel{\iota}{\le}(V_{1},V_{3})$, $V_{4}\stackrel{\iota}{\le}V_{3}$,
and $V_{3}\stackrel{\iota}{\le}V_{4}$. The red zig-zag edges are
Network 1. The black solid edges are Network 2.}

\end{figure}

Refer to Figure \ref{fig:networks} for an illustration of the networks.
Let
\[
U:=(U_{1},\ldots,U_{n}),
\]
\[
T_{i}:=\mathrm{S}\big(((X_{i}^{j})_{j=1}^{k},Y_{i},Z_{i});\,U\big).
\]
Our goal is to show that the networks imply that $T_{1},\ldots,T_{n}$
satisfy the antecedents in the implication problem in Corollary \ref{cor:undec_eq}.
We first show that $\mathrm{S}(X_{i}^{j};U)\stackrel{\iota}{=}T_{i}$
for all $j$. Intuitively, the variables $X_{i}^{1},\ldots,X_{i}^{k}$
act as ``copies'' of the sufficient statistic $T_{i}$, so that
we can impose separate functional dependencies on the copies without
them interfering each other.

\medskip{}

\begin{lem}
Networks 1 and 2 imply that 
\[
\mathrm{S}(Y_{i};U)\stackrel{\iota}{=}\mathrm{S}(Z_{i};U)\stackrel{\iota}{=}\mathrm{S}(X_{i}^{j};U)\stackrel{\iota}{=}T_{i}
\]
for every $i,j$.
\end{lem}
\begin{IEEEproof}
Since Network 1 implies that $U\to Y_{i}\to X_{i}^{1}\to\cdots\to X_{i}^{k}\to Z_{i}$
forms a Markov chain, by Lemma \ref{lem:sufficient_add}, $\mathrm{S}(Y_{i};U)\stackrel{\iota}{=}T_{i}$.
Also, the Markov chain $U\to Y_{i}\to Z_{i}$ gives $\mathrm{S}((Y_{i},Z_{i});U)\stackrel{\iota}{=}\mathrm{S}(Y_{i};U)\stackrel{\iota}{=}T_{i}$.
Since Network 2 implies that $U\to Z_{i}\to Y_{i}$ forms a Markov
chain, by Lemma \ref{lem:sufficient_add}, $\mathrm{S}(Z_{i};U)\stackrel{\iota}{=}\mathrm{S}((Y_{i},Z_{i});U)\stackrel{\iota}{=}T_{i}$.
Network 1 gives $U\to Y_{i}\to X_{i}^{j}\to Z_{i}$, and hence by
Lemma \ref{lem:sufficient_insert}, we have the Markov chain 
\[
U\to T_{i}\to Y_{i}\to X_{i}^{j}\to Z_{i}\to T_{i}
\]
by definition of the sufficient statistics $\mathrm{S}(Y_{i};U)\stackrel{\iota}{=}\mathrm{S}(Z_{i};U)\stackrel{\iota}{=}T_{i}$.
Since $U\to T_{i}\to Y_{i}$, again by Lemma \ref{lem:sufficient_add},
$\mathrm{S}(T_{i};U)\stackrel{\iota}{=}\mathrm{S}((T_{i},Y_{i});U)\stackrel{\iota}{=}\mathrm{S}(Y_{i};U)\stackrel{\iota}{=}T_{i}$
since $(T_{i},Y_{i})\stackrel{\iota}{=}Y_{i}$. We have $T_{i}\to X_{i}^{j}\to T_{i}$,
and hence $X_{i}^{j}\stackrel{\iota}{\ge}T_{i}$. Since $U\to T_{i}\to X_{i}^{j}$,
by Lemma \ref{lem:sufficient_add}, $T_{i}\stackrel{\iota}{=}\mathrm{S}(T_{i};U)\stackrel{\iota}{=}\mathrm{S}((T_{i},X_{i}^{j});U)\stackrel{\iota}{=}\mathrm{S}(X_{i}^{j};U)$. 
\end{IEEEproof}
\medskip{}

We show that $T_{1},\ldots,T_{n}$ must satisfy the mutual independence
conditions in Corollary \ref{cor:undec_eq}. 

\medskip{}

\begin{lem}
\label{lem:networks_mi}Networks 1 and 2 imply that $(T_{i})_{i\in\mathcal{C}_{1}}$
are mutually independent, and $(T_{i})_{i\in\mathcal{C}_{2}}$ are
mutually independent.
\end{lem}
\begin{IEEEproof}
Since there is no edge pointing towards $(U_{i})_{i\in\mathcal{C}_{2}}$
in Network 2, it imposes that $(U_{i})_{i\in\mathcal{C}_{2}}$ are
mutually independent. Fix any $i\in\mathcal{C}_{2}$. Network 1 implies
the Markov chain $Y_{\mathcal{C}_{2}\backslash\{i\}}\to U_{\mathcal{C}_{2}\backslash\{i\}}\to U_{i}\to Y_{i}$
by $d$-separation. Considering that $U_{\mathcal{C}_{2}\backslash\{i\}}\perp\!\!\!\perp U_{i}$,
we have $Y_{\mathcal{C}_{2}\backslash\{i\}}\perp\!\!\!\perp Y_{i}$,
and hence $T_{\mathcal{C}_{2}\backslash\{i\}}\perp\!\!\!\perp T_{i}$
since $T_{i}\stackrel{\iota}{=}\mathrm{S}(Y_{i};U)\stackrel{\iota}{\le}Y_{i}$.
The same arguments hold for $(T_{i})_{i\in\mathcal{C}_{1}}$.
\end{IEEEproof}
\medskip{}

To show that $T_{1},\ldots,T_{n}$ satisfy functional dependency conditions
in Corollary \ref{cor:undec_eq}, we require the following tool. 

\medskip{}

\begin{lem}
\label{lem:networks_group}For any set $\mathcal{S}\subseteq\{1,\ldots,n\}$
and $j\in\{1,\ldots,k\}$, Networks 1 and 2 imply that $T_{\mathcal{S}}$
is a sufficient statistic of $X_{\mathcal{S}}^{j}$ (or $Y_{\mathcal{S}}$,
or $Z_{\mathcal{S}}$) for $U$.
\end{lem}
\begin{IEEEproof}
Network 1 implies the Bayesian network with edges $U\to X_{i}^{j}$
for $i=1,\ldots,n$, which implies the Bayesian network with edges
$U\to T_{i}\to X_{i}^{j}$ for $i=1,\ldots,n$ by invoking Lemma \ref{lem:sufficient_insert}
to insert the sufficient statistics $T_{i}\stackrel{\iota}{=}\mathrm{S}(X_{i}^{j};U)$.
By $d$-separation, we have the Markov chain $U\to T_{\mathcal{S}}\to X_{\mathcal{S}}^{j}$.
Hence $T_{\mathcal{S}}$ is a sufficient statistic of $X_{\mathcal{S}}^{j}$
for $U$. The same arguments hold for $Y_{\mathcal{S}}$ and $Z_{\mathcal{S}}$.

\end{IEEEproof}
\medskip{}

We can now show that $T_{1},\ldots,T_{n}$ must satisfy the functional
dependency conditions in Corollary \ref{cor:undec_eq}. 

\medskip{}

\begin{lem}
\label{lem:networks_fd}Networks 1 and 2 imply that $T_{b_{j}}\stackrel{\iota}{\le}T_{\mathcal{A}_{j}}$
for every $j=1,\ldots,k$.
\end{lem}
\begin{IEEEproof}
Fix any $j\in\{1,\ldots,k\}$. The only edges incident with $X_{b_{j}}^{j}$
in Network 2 are $X_{i}^{j}\to X_{b_{j}}^{j}$ for $i\in\mathcal{A}_{j}$.
Therefore, $X_{b_{j}}^{j}$ is conditionally independent of every
other nodes given $X_{\mathcal{A}_{j}}^{j}$. In particular, we have
the Markov chain $Y_{b_{j}}\to X_{\mathcal{A}_{j}}^{j}\to X_{b_{j}}^{j}$.
Since $T_{b_{j}}\stackrel{\iota}{\le}Y_{b_{j}}$ and $T_{b_{j}}\stackrel{\iota}{\le}X_{b_{j}}^{j}$,
we have $T_{b_{j}}\to X_{\mathcal{A}_{j}}^{j}\to T_{b_{j}}$, and
hence $T_{b_{j}}\stackrel{\iota}{\le}X_{\mathcal{A}_{j}}^{j}$. From
Network 1, we have $Y_{b_{j}}\to U\to X_{\mathcal{A}_{j}}^{j}$ by
$d$-separation. Since Lemma \ref{lem:networks_group} shows that
$T_{\mathcal{A}_{j}}$ is a sufficient statistic of $X_{\mathcal{A}_{j}}^{j}$
for $U$, invoking Lemma \ref{lem:sufficient_insert}, we have $Y_{b_{j}}\to U\to T_{\mathcal{A}_{j}}\to X_{\mathcal{A}_{j}}^{j}$.
Hence, we have the Markov chain 
\[
T_{b_{j}}\to Y_{b_{j}}\to U\to T_{\mathcal{A}_{j}}\to X_{\mathcal{A}_{j}}^{j}\to T_{b_{j}},
\]
implying that $T_{b_{j}}\stackrel{\iota}{\le}T_{\mathcal{A}_{j}}$.
\end{IEEEproof}
\medskip{}

We now complete the proof of Theorem \ref{thm:two_bayes} by a reduction
from the problem in Corollary \ref{cor:undec_eq}. We will also see
the purpose of introducing $b_{0}'$.

\medskip{}

\begin{thm}
Networks 1 and 2 imply the conditional independency $Y_{b_{0}}\perp\!\!\!\perp Y_{b_{0}'}|Y_{\mathcal{A}_{0}}$
if and only if the implication in Corollary \ref{cor:undec_eq} holds.
\end{thm}
\begin{IEEEproof}
First prove the ``if'' direction. Assume the implication in Corollary
\ref{cor:undec_eq} holds, and Networks 1 and 2 are satisfied. By
Lemma \ref{lem:networks_mi}, $(T_{i})_{i\in\mathcal{C}_{1}}$ are
mutually independent, and $(T_{i})_{i\in\mathcal{C}_{2}}$ are mutually
independent. By Lemma \ref{lem:networks_fd}, $T_{b_{j}}\stackrel{\iota}{\le}T_{\mathcal{A}_{j}}$
for every $j=1,\ldots,k$. Hence, the antecedents in the the implication
in Corollary \ref{cor:undec_eq} are satisfied, and we have $T_{b_{0}}\stackrel{\iota}{=}T_{b_{0}'}\stackrel{\iota}{\le}T_{\mathcal{A}_{0}}$.
Network 1 implies the Bayesian network 
\[
\begin{array}{ccccc}
 &  & U\\
 & \swarrow & \downarrow & \searrow\\
Y_{b_{0}} &  & Y_{\mathcal{A}_{0}} &  & Y_{b_{0}'}
\end{array}
\]
which implies the Bayesian network 
\[
\begin{array}{ccccc}
 &  & U\\
 & \swarrow & \downarrow & \searrow\\
T_{b_{0}} &  & T_{\mathcal{A}_{0}} &  & T_{b_{0}'}\\
\downarrow &  & \downarrow &  & \downarrow\\
Y_{b_{0}} &  & Y_{\mathcal{A}_{0}} &  & Y_{b_{0}'}
\end{array}
\]
by invoking Lemma \ref{lem:sufficient_insert} and \ref{lem:networks_group}
to insert the sufficient statistics. By $d$-separation, we have $Y_{b_{0}}\perp\!\!\!\perp Y_{b_{0}'}|(T_{b_{0}},T_{b_{0}'},T_{\mathcal{A}_{0}},Y_{\mathcal{A}_{0}})$.
Since $T_{b_{0}}\stackrel{\iota}{=}T_{b_{0}'}\stackrel{\iota}{\le}T_{\mathcal{A}_{0}}\stackrel{\iota}{\le}Y_{\mathcal{A}_{0}}$,
this gives $Y_{b_{0}}\perp\!\!\!\perp Y_{b_{0}'}|Y_{\mathcal{A}_{0}}$.
Hence, Networks 1 and 2 imply $Y_{b_{0}}\perp\!\!\!\perp Y_{b_{0}'}|Y_{\mathcal{A}_{0}}$.\footnote{This argument would not work if $b_{0}'=b_{0}$. This is the reason
we require two identical copies $V_{b_{0}}\stackrel{\iota}{=}V_{b_{0}'}$
in Corollary \ref{cor:undec_eq}.}

Then prove the ``only if'' direction. Assume Networks 1 and 2 imply
the conditional independency $Y_{b_{0}}\perp\!\!\!\perp Y_{b_{0}'}|Y_{\mathcal{A}_{0}}$,
and assume random variables $V_{1},\ldots,V_{n}$ satisfy the antecedents
in the implication in Corollary \ref{cor:undec_eq}, i.e., $(V_{i})_{i\in\mathcal{C}_{1}}$
are mutually independent, $(V_{i})_{i\in\mathcal{C}_{2}}$ are mutually
independent, and $V_{b_{j}}\stackrel{\iota}{\le}V_{\mathcal{A}_{j}}$
for $j=1,\ldots,k$. Take $U_{i}=X_{i}^{j}=Y_{i}=Z_{i}=V_{i}$ for
$i=1,\ldots,n$, $j=1,\ldots,k$. Network 1 is satisfied since $(U_{i})_{i\in\mathcal{C}_{1}}$
are mutually independent, and $U_{i}=Y_{i}=X_{i}^{j}=Z_{i}$. For
Network 2, since $X_{b_{j}}^{j}=V_{b_{j}}\stackrel{\iota}{\le}V_{\mathcal{A}_{j}}=X_{\mathcal{A}_{j}}^{j}$,
the Markov condition for $X_{b_{j}}^{j}$ is satisfied. The Markov
conditions for other $X_{i}^{j}$ are satisfied since $X_{i}^{j}=U_{i}$.
It is straightforward to check that all other conditions are satisfied.
Hence Networks 1 and 2 are satisfied, and we have $Y_{b_{0}}\perp\!\!\!\perp Y_{b_{0}'}|Y_{\mathcal{A}_{0}}$,
or equivalently, $V_{b_{0}}\perp\!\!\!\perp V_{b_{0}'}|V_{\mathcal{A}_{0}}$.
Recall that Corollary \ref{cor:undec_eq} also imposes that $V_{b_{0}}\stackrel{\iota}{=}V_{b_{0}'}$.
Hence, $V_{b_{0}}\stackrel{\iota}{\le}V_{\mathcal{A}_{0}}$, and the
implication in Corollary \ref{cor:undec_eq} holds.
\end{IEEEproof}
\medskip{}

As a result, the problem in Theorem \ref{thm:two_bayes} is undecidable
by a reduction from the problem in Corollary \ref{cor:undec_eq}.
Note that $\mathcal{A}_{0}$ is a singleton set, so $Y_{b_{0}}\perp\!\!\!\perp Y_{b_{0}'}|Y_{\mathcal{A}_{0}}$
only involves three single random variables.

\medskip{}

\section{Conclusion and Future Directions}

In this paper, we showed that two Bayesian network structures have
undecidable conditional independencies (Theorem \ref{thm:two_bayes}).
Also, the implication problem of Bayesian network structures is undecidable
when there are at least two networks in the antecedent (Corollary
\ref{cor:bayes_impl}). This is in stark contrast to the situation
with one Bayesian network structure, where the results on $d$-separation
shows that the conditional independencies of one Bayesian network,
and the implication problem of Bayesian network structures when there
is one network in the antecedent, are decidable \cite{verma1990causal,geiger1990logic,geiger1990d}.
Our undecidability results were proved using some techniques in the
proof of the undecidability of conditional independence implication
in \cite{li2023undecidability}, together with some novel constructions
to impose functional dependency conditions on sufficient statistics.
It is perhaps interesting that, although we cannot use one Bayesian
network to impose functional dependencies among random variables,
we can do so with two Bayesian networks, but the functional dependencies
are not imposed on the random variables themselves, but on their sufficient
statistics.

We now discuss several related problems, where it is still unknown
whether they are decidable or not.

\smallskip{}

\begin{itemize}
\item \textbf{Constructing a Bayesian network from a set of conditional
independencies.} Is there an algorithm that takes a set of CIs $\mathcal{S}\subseteq(2^{\{1,\ldots,n\}})^{3}$
among random variables $X_{1},\ldots,X_{n}$ as input, and outputs
a Bayesian network structure $G$ which is logically equivalent to
$\mathcal{S}$, i.e., they imply the same set of CIs
\[
\mathrm{Cl}(\mathcal{S})=\mathrm{Cl}(\mathcal{I}_{\ell}(G)),
\]
or output $\emptyset$ if no such $G$ exists? Alternatively, is it
decidable to determine whether a given set of CIs $\mathcal{S}$ and
a given Bayesian network structure  $G$ are logically equivalent?
While this appears to be a fundamental problem about Bayesian network,
its decidability seems to be unknown. Note that if we relax the logical
equivalence to implication (i.e., change the ``$=$'' in the above
formula to ``$\subseteq$'' or ``$\supseteq$''), then it is decidable
to determine whether $\mathrm{Cl}(\mathcal{S})\subseteq\mathrm{Cl}(\mathcal{I}_{\ell}(G))$,
i.e., whether $G$ implies $\mathcal{S}$ ($d$-separation \cite{verma1990causal,geiger1990logic,geiger1990d}),
and undecidable to determine whether $\mathrm{Cl}(\mathcal{S})\supseteq\mathrm{Cl}(\mathcal{I}_{\ell}(G))$,
i.e., whether $G$ is implied by $\mathcal{S}$ (same as CI implication
\cite{li2023undecidability}). \smallskip{}
\item \textbf{Compatibility of Bayesian networks.} Given Bayesian network
structures $G_{1},\ldots,G_{k}$, determine whether there exist random
variables $X_{1},\ldots,X_{n}$ that are not all mutually independent
that simultaneously satisfy all of $G_{1},\ldots,G_{k}$. Equivalently,
determine whether $\mathcal{I}(G_{1},\ldots,G_{k})$ (the CIs implied
by $G_{1},\ldots,G_{k}$) contains every disjoint $3$-tuple $(\mathcal{A},\mathcal{B},\mathcal{C})$.
If $\mathcal{I}(G_{1},\ldots,G_{k})$ contains every disjoint $3$-tuple,
then $G_{1},\ldots,G_{k}$ are ``incompatible'' in the sense that
they cannot be satisfied simultaneously, except for the trivial case
where $X_{1},\ldots,X_{n}$ are mutually independent.\smallskip{}
\item \textbf{Conditional independencies of a collection of Markov chains.}
Given $a_{i}^{j}\in\{1,\ldots,n\}$ for $j\in\{0,\ldots,k\}$, $i\in\{1,\ldots,\ell_{j}\}$,
determine whether the collection of Markov chains $X_{a_{1}^{j}}\to X_{a_{2}^{j}}\to\cdots\to X_{a_{\ell_{j}}^{j}}$
for $j=1,\ldots,k$ about the random variables $X_{1},\ldots,X_{n}$
imply the Markov chain $X_{a_{0}^{j}}\to X_{a_{0}^{j}}\to\cdots\to X_{a_{\ell_{0}}^{j}}$.
This problem is clearly decidable if $k$ is fixed to $1$ (since
a Markov chain is a Bayesian network, this can be solved by $d$-separation),
and clearly undecidable if $k$ can be arbitrarily chosen (same as
CI implication \cite{li2023undecidability}). Is it decidable when
$k$ is fixed to $2$? What is the largest fixed value of $k$ where
this is decidable?\smallskip{}
\end{itemize}
\medskip{}

\section{Acknowledgement}

This work was partially supported by two grants from the Research
Grants Council of the Hong Kong Special Administrative Region, China
{[}Project No.s: CUHK 24205621 (ECS), CUHK 14209823 (GRF){]}. The
diagrams in this paper are created using Quiver \cite{Arkor_quiver_2024}
or Ipe \cite{ipe_2023}.

\medskip{}

\appendix{}

\subsection{Proof of Theorem \ref{thm:three_undec}\label{subsec:pf_three_undec}}

We now examine the construction of the undecidable family of conditional
independence implication problems in \cite[Theorem 1]{li2023undecidability}
(we will not go into the details of the proof in \cite{li2023undecidability}
since it is quite complicated). In \cite{li2023undecidability}, the
conditional independencies are constructed using the predicates $\mathrm{tri}$,
$\mathrm{fnf}$, $\mathrm{ueq}$, $\mathrm{end}_{i,j}$, $\mathrm{conv}_{i,j}^{k,l}$
and $\mathrm{comp}_{i,j}$. All random variables that appear in $\mathrm{tri}$,
$\mathrm{fnf}$, $\mathrm{end}_{i,j}$, $\mathrm{conv}_{i,j}^{k,l}$
and $\mathrm{comp}_{i,j}$ are restricted  to be functions of the
three independent random variables $A_{1},A_{2},A_{3}$ in $\mathrm{fnf}$.
Also, in \cite{li2023undecidability}, all random variables are restricted
to be uniformly distributed with the same cardinality using $\mathrm{fnf}$
and $\mathrm{ueq}$. Only the functional dependency relation $\stackrel{\iota}{\ge}$
and the unconditional independence relation in the form ``$X\perp\!\!\!\perp Y$''
(where $X$ and $Y$ are single random variables, not joint random
variables; this only appear in $\mathrm{tri}$) and ``$A_{1}\perp\!\!\!\perp A_{2}\perp\!\!\!\perp A_{3}$''
(appears in $\mathrm{fnf}$) appear in the definitions of the predicates. 

Therefore, \cite{li2023undecidability} establishes the undecidability
of the special form of conditional independence implication problems,
where all random variables are uniformly distributed with the same
cardinality, and are functions of three independent random variables
$A_{1},A_{2},A_{3}$; and we only impose functional dependency relations
$\stackrel{\iota}{\ge}$ and additional unconditional independence
relations in the form ``$X\perp\!\!\!\perp Y$'' among random variables.

\medskip{}

\bibliographystyle{IEEEtran}
\bibliography{ref}

\begin{thebibliography}{10}
\providecommand{\url}[1]{#1}
\csname url@samestyle\endcsname
\providecommand{\newblock}{\relax}
\providecommand{\bibinfo}[2]{#2}
\providecommand{\BIBentrySTDinterwordspacing}{\spaceskip=0pt\relax}
\providecommand{\BIBentryALTinterwordstretchfactor}{4}
\providecommand{\BIBentryALTinterwordspacing}{\spaceskip=\fontdimen2\font plus
\BIBentryALTinterwordstretchfactor\fontdimen3\font minus
  \fontdimen4\font\relax}
\providecommand{\BIBforeignlanguage}[2]{{%
\expandafter\ifx\csname l@#1\endcsname\relax
\typeout{** WARNING: IEEEtran.bst: No hyphenation pattern has been}%
\typeout{** loaded for the language `#1'. Using the pattern for}%
\typeout{** the default language instead.}%
\else
\language=\csname l@#1\endcsname
\fi
#2}}
\providecommand{\BIBdecl}{\relax}
\BIBdecl

\bibitem{pearl1988probabilistic}
J.~Pearl, \emph{Probabilistic Reasoning in Intelligent Systems}.\hskip 1em plus
  0.5em minus 0.4em\relax Morgan Kaufmann, 1988.

\bibitem{verma1990causal}
T.~Verma and J.~Pearl, ``Causal networks: Semantics and expressiveness,'' in
  \emph{Machine intelligence and pattern recognition}.\hskip 1em plus 0.5em
  minus 0.4em\relax Elsevier, 1990, vol.~9, pp. 69--76.

\bibitem{geiger1990logic}
D.~Geiger and J.~Pearl, ``On the logic of causal models,'' in \emph{Machine
  intelligence and pattern recognition}.\hskip 1em plus 0.5em minus 0.4em\relax
  Elsevier, 1990, vol.~9, pp. 3--14.

\bibitem{geiger1990d}
D.~Geiger, T.~Verma, and J.~Pearl, ``d-separation: From theorems to
  algorithms,'' in \emph{Machine intelligence and pattern recognition}.\hskip
  1em plus 0.5em minus 0.4em\relax Elsevier, 1990, vol.~10, pp. 139--148.

\bibitem{matzkevich1992topological}
I.~Matzkevich and B.~Abramson, ``The topological fusion of bayes nets,'' in
  \emph{Uncertainty in Artificial Intelligence}.\hskip 1em plus 0.5em minus
  0.4em\relax Elsevier, 1992, pp. 191--198.

\bibitem{matzkevich1993some}
------, ``Some complexity considerations in the combination of belief
  networks,'' in \emph{Uncertainty in Artificial Intelligence}.\hskip 1em plus
  0.5em minus 0.4em\relax Elsevier, 1993, pp. 152--158.

\bibitem{xiang1998verification}
Y.~Xiang, ``Verification of {DAG} structures in cooperative belief
  network-based multiagent systems,'' \emph{Networks: An International
  Journal}, vol.~31, no.~3, pp. 183--191, 1998.

\bibitem{pennock1999graphical}
D.~M. Pennock and M.~P. Wellman, ``Graphical representations of consensus
  belief,'' in \emph{Proceedings of the Fifteenth conference on Uncertainty in
  artificial intelligence}, 1999, pp. 531--540.

\bibitem{wong2001constructing}
S.~K.~M. Wong and C.~J. Butz, ``Constructing the dependency structure of a
  multiagent probabilistic network,'' \emph{IEEE Transactions on Knowledge and
  Data Engineering}, vol.~13, no.~3, pp. 395--415, 2001.

\bibitem{ron1997combining}
E.~C. Ron, J.~M.~G. Llorente, and A.~S. Hadi, ``Combining multiple directed
  graphical representations into a single probabilistic model,'' in
  \emph{CAEPIA'97: actas}.\hskip 1em plus 0.5em minus 0.4em\relax Vicent Botti,
  1997, pp. 645--652.

\bibitem{del2003qualitative}
J.~Del~Sagrado and S.~Moral, ``Qualitative combination of bayesian networks,''
  \emph{International Journal of Intelligent Systems}, vol.~18, no.~2, pp.
  237--249, 2003.

\bibitem{jiang2005pgmc}
C.-a. Jiang, T.-Y. Leong, and P.~Kim-Leng, ``{PGMC}: a framework for
  probabilistic graphical model combination,'' in \emph{AMIA Annual Symposium
  Proceedings}, vol. 2005.\hskip 1em plus 0.5em minus 0.4em\relax American
  Medical Informatics Association, 2005, p. 370.

\bibitem{li2008recovering}
W.~Li, W.~Liu, and K.~Yue, ``Recovering the global structure from multiple
  local bayesian networks,'' \emph{International Journal on Artificial
  Intelligence Tools}, vol.~17, no.~06, pp. 1067--1088, 2008.

\bibitem{feng2014novel}
G.~Feng, J.-D. Zhang, and S.~S. Liao, ``A novel method for combining {B}ayesian
  networks, theoretical analysis, and its applications,'' \emph{Pattern
  Recognition}, vol.~47, no.~5, pp. 2057--2069, 2014.

\bibitem{tabar2018novel}
V.~R. Tabar and F.~Elahi, ``A novel method for aggregation of {B}ayesian
  networks without considering an ancestral ordering,'' \emph{Applied
  Artificial Intelligence}, vol.~32, no.~2, pp. 214--227, 2018.

\bibitem{pearl1987graphoids}
J.~Pearl and A.~Paz, ``Graphoids: a graph-based logic for reasoning about
  relevance relations,'' \emph{Advances in Artificial Intelligence}, pp.
  357--363, 1987.

\bibitem{studeny1989multiinformation}
M.~Studen{\'{y}}, ``Multiinformation and the problem of characterization of
  conditional independence relations,'' \emph{Problems of Control and
  Information Theory}, vol.~18, pp. 3--16, 1989.

\bibitem{studeny1990conditional}
------, ``Conditional independence relations have no finite complete
  characterization,'' \emph{Information Theory, Statistical Decision Functions
  and Random Processes}, pp. 377--396, 1992.

\bibitem{malvestuto1992unique}
F.~M. Malvestuto, ``A unique formal system for binary decompositions of
  database relations, probability distributions, and graphs,''
  \emph{Information Sciences}, vol.~59, no. 1-2, pp. 21--52, 1992.

\bibitem{geiger1993logical}
D.~Geiger and J.~Pearl, ``Logical and algorithmic properties of conditional
  independence and graphical models,'' \emph{The Annals of Statistics}, pp.
  2001--2021, 1993.

\bibitem{geiger1987non}
D.~Geiger, \emph{The non-axiomatizability of dependencies in directed acyclic
  graphs}.\hskip 1em plus 0.5em minus 0.4em\relax Technical report R-83,
  University of California (Los Angeles), Computer Science Department, 1987.

\bibitem{li2008causal}
S.~Li, ``Causal models have no complete axiomatic characterization,''
  \emph{arXiv preprint arXiv:0804.2401}, 2008.

\bibitem{li2023undecidability}
C.~T. Li, ``Undecidability of network coding, conditional information
  inequalities, and conditional independence implication,'' \emph{IEEE
  Transactions on Information Theory}, vol.~69, no.~6, pp. 3493--3510, 2023.

\bibitem{kuhne2022entropic}
L.~K{\"u}hne and G.~Yashfe, ``On entropic and almost multilinear
  representability of matroids,'' \emph{arXiv preprint arXiv:2206.03465}, Jun
  2022.

\bibitem{herrmann1995undecidability}
C.~Herrmann, ``On the undecidability of implications between embedded
  multivalued database dependencies,'' \emph{Information and Computation}, vol.
  122, no.~2, pp. 221--235, 1995.

\bibitem{kuhne2019representability}
L.~K{\"u}hne and G.~Yashfe, ``Representability of matroids by c-arrangements is
  undecidable,'' \emph{arXiv preprint arXiv:1912.06123}, 2019.

\bibitem{li2022undecidabilityaffine}
C.~T. Li, ``The undecidability of conditional affine information inequalities
  and conditional independence implication with a binary constraint,''
  \emph{IEEE Transactions on Information Theory}, vol.~68, no.~12, pp.
  7685--7701, 2022.

\bibitem{li2022netcodeundec}
------, ``The undecidability of network coding with some fixed-size messages
  and edges,'' in \emph{2022 IEEE International Symposium on Information Theory
  (ISIT)}.\hskip 1em plus 0.5em minus 0.4em\relax IEEE, 2022, pp. 31--36.

\bibitem{matuvs1995conditional}
F.~Mat{\'u}{\v{s}} and M.~Studen{\`y}, ``Conditional independences among four
  random variables {I},'' \emph{Combinatorics, Probability and Computing},
  vol.~4, no.~3, pp. 269--278, 1995.

\bibitem{yeung1997framework}
R.~W. Yeung, ``A framework for linear information inequalities,'' \emph{IEEE
  Trans. Inf. Theory}, vol.~43, no.~6, pp. 1924--1934, 1997.

\bibitem{yeung2021machine}
R.~W. Yeung and C.~T. Li, ``Machine-proving of entropy inequalities,''
  \emph{IEEE BITS the Information Theory Magazine}, vol.~1, no.~1, pp. 12--22,
  September 2021.

\bibitem{li2023automated}
C.~T. Li, ``An automated theorem proving framework for information-theoretic
  results,'' \emph{IEEE Transactions on Information Theory}, vol.~69, no.~11,
  pp. 6857--6877, 2023.

\bibitem{boege2024self}
T.~Boege, J.~H. Bolt, and M.~Studen{\`y}, ``Self-adhesivity in lattices of
  abstract conditional independence models,'' \emph{arXiv preprint
  arXiv:2402.14053}, 2024.

\bibitem{yedidia2016relatively}
A.~Yedidia and S.~Aaronson, ``A relatively small {T}uring machine whose
  behavior is independent of set theory,'' \emph{arXiv preprint
  arXiv:1605.04343}, 2016.

\bibitem{michel2009busy}
P.~Michel, ``The busy beaver competition: a historical survey,'' \emph{arXiv
  preprint arXiv:0906.3749}, 2009.

\bibitem{gurevich1966problem}
Y.~S. Gurevich, ``The problem of equality of words for certain classes of
  semigroups,'' \emph{Algebra i logika}, vol.~5, no.~5, pp. 25--35, 1966.

\bibitem{cooper1990computational}
G.~F. Cooper, ``The computational complexity of probabilistic inference using
  {B}ayesian belief networks,'' \emph{Artificial intelligence}, vol.~42, no.
  2-3, pp. 393--405, 1990.

\bibitem{dagum1993approximating}
P.~Dagum and M.~Luby, ``Approximating probabilistic inference in {B}ayesian
  belief networks is {NP}-hard,'' \emph{Artificial intelligence}, vol.~60,
  no.~1, pp. 141--153, 1993.

\bibitem{roth1996hardness}
D.~Roth, ``On the hardness of approximate reasoning,'' \emph{Artificial
  Intelligence}, vol.~82, no. 1-2, pp. 273--302, 1996.

\bibitem{armstrong1974dependency}
W.~W. Armstrong, ``Dependency structures of data base relationships,'' in
  \emph{IFIP congress}, vol.~74.\hskip 1em plus 0.5em minus 0.4em\relax Geneva,
  Switzerland, 1974, pp. 580--583.

\bibitem{li2021first}
C.~T. Li, ``First-order theory of probabilistic independence and single-letter
  characterizations of capacity regions,'' \emph{IEEE Transactions on
  Information Theory}, vol.~69, no.~12, pp. 7584--7601, 2023.

\bibitem{koller2009probabilistic}
D.~Koller and N.~Friedman, \emph{Probabilistic graphical models: principles and
  techniques}.\hskip 1em plus 0.5em minus 0.4em\relax MIT press, 2009.

\bibitem{theodoridis2020machine}
S.~Theodoridis, \emph{Machine Learning: A {B}ayesian and Optimization
  Perspective}.\hskip 1em plus 0.5em minus 0.4em\relax Academic Press, 2020.

\bibitem{meek1995strong}
C.~Meek, ``Strong completeness and faithfulness in bayesian networks,'' in
  \emph{Proc. Conf. on Uncertainty in Artificial Intelligence (UAI-95)}, 1995,
  pp. 411--418.

\bibitem{meek1997graphical}
------, ``Graphical models: Selecting causal and statistical models,'' Ph.D.
  dissertation, Carnegie Mellon University, 1997.

\bibitem{kocka2013characterizing}
T.~Kocka, R.~R. Bouckaert, and M.~Studen{\`y}, ``On characterizing inclusion of
  bayesian networks,'' \emph{arXiv preprint arXiv:1301.2282}, 2013.

\bibitem{dawid1979conditional}
A.~P. Dawid, ``Conditional independence in statistical theory,'' \emph{Journal
  of the Royal Statistical Society: Series B (Methodological)}, vol.~41, no.~1,
  pp. 1--15, 1979.

\bibitem{marshall1979inequalities}
A.~W. Marshall, I.~Olkin, and B.~C. Arnold, \emph{Inequalities: theory of
  majorization and its applications}.\hskip 1em plus 0.5em minus 0.4em\relax
  New York, Dordrecht, Heidelberg, London: Springer, 2011.

\bibitem{Arkor_quiver_2024}
\BIBentryALTinterwordspacing
N.~Arkor, ``{quiver},'' \url{https://github.com/varkor/quiver}, Apr. 2024.
  [Online]. Available: \url{https://github.com/varkor/quiver}
\BIBentrySTDinterwordspacing

\bibitem{ipe_2023}
\BIBentryALTinterwordspacing
O.~Cheong, ``{Ipe},'' \url{https://github.com/otfried/ipe}, 2023. [Online].
  Available: \url{https://github.com/otfried/ipe}
\BIBentrySTDinterwordspacing

\end{thebibliography}

\end{document}